\documentclass[a4paper,twocolumn,10pt,accepted=2024-02-08]{quantumarticle}
\pdfoutput=1
\usepackage[utf8]{inputenc}
\usepackage[english]{babel}
\usepackage[T1]{fontenc}

\usepackage{graphicx}
\usepackage{setspace}
\usepackage{amsmath,amssymb,amsthm}
\usepackage{bbold}
\usepackage{hyperref}
\usepackage{xcolor}
\usepackage{enumitem}
\usepackage[utf8]{inputenc}
\usepackage{cancel} 
\usepackage{qcircuit}
\usepackage[ruled,vlined]{algorithm2e}
\usepackage[version=4]{mhchem}
\usepackage{siunitx}

\usepackage[numbers,compress]{natbib}

\newcommand{\bra}[1]{\ensuremath{\langle{#1}\rvert}}
\newcommand{\ket}[1]{\ensuremath{\lvert{#1}\rangle}}
\newcommand{\braket}[2]{\ensuremath{\langle{#1}\rvert{#2}\rangle}}

\renewcommand{\Re}{\mathrm{Re}}

\DeclareMathOperator{\Var}{Var}

\newcommand{\R}{{\mathbf{R}}} 
\newcommand{\C}{{\mathcal{C}}} 
\newcommand{\bftheta}{{\boldsymbol{\theta}}}
\newcommand{\bfkappa}{{\boldsymbol{\kappa}}}

\newcommand{\eigstate}{\Phi} 
\newcommand{\gaugedstate}{\Psi} 
\newcommand{\thetaex}{\bftheta^*} 
\newcommand{\thetaest}{\tilde{\bftheta}} 

\newtheorem{lemma}{Lemma}
\newtheorem{theorem}{Theorem}

\newcommand{\Google}{\affiliation{%
Google Research, Munich, 80636 Bavaria, Germany}}

\newcommand{\Leiden}{\affiliation{%
Instituut-Lorentz, Universiteit Leiden, 2300RA Leiden, The Netherlands}}

\newcommand{\ICFO}{\affiliation{%
ICFO - Institut de Ciències Fotòniques, 08860 Castelldefels (Barcelona), Spain}}

\newcommand{\VU}{\affiliation{%
Theoretical Chemistry, Vrije Universiteit, 1081HV Amsterdam,
The Netherlands}}

\newcommand{\PASQAL}{\affiliation{
PASQAL SAS, 2 av. Augustin Fresnel Palaiseau, 91120, France}}

\begin{document}
	
\title{A hybrid quantum algorithm to detect conical intersections}

\author{Emiel Koridon}
\Leiden \VU

\author{Joana Fraxanet}
\ICFO

\author{Alexandre Dauphin}
\ICFO
\PASQAL

\author{Lucas Visscher}
\VU

\author{Thomas E. O'Brien}
\Google \Leiden

\author{Stefano Polla}
\email{polla@lorentz.leidenuniv.nl}
\Google \Leiden

\begin{abstract}
    Conical intersections are topologically protected crossings between the potential energy surfaces of a molecular Hamiltonian, known to play an important role in chemical processes such as photoisomerization and non-radiative relaxation. 
    They are characterized by a non-zero Berry phase, which is a topological invariant defined on a closed path in atomic coordinate space, taking the value $\pi$ when the path encircles the intersection manifold. 
    In this work, we show that for real molecular Hamiltonians, the Berry phase can be obtained by tracing a local optimum of a variational ansatz along the chosen path and estimating the overlap between the initial and final state with a control-free Hadamard test.
    Moreover, by discretizing the path into $N$ points, we can use $N$ single Newton-Raphson steps to update our state non-variationally.
    Finally, since the Berry phase can only take two discrete values (0 or $\pi$), our procedure succeeds even for a cumulative error bounded by a constant; 
    this allows us to bound the total sampling cost and to readily verify the success of the procedure.
    We demonstrate numerically the application of our algorithm on small toy models of the formaldimine molecule (\ce{H2C=NH}).
\end{abstract}

\maketitle

\tableofcontents

\section{Introduction}
\label{sec:introduction}

Conical intersections (CI) are degeneracy points in the Born-Oppenheimer molecular structure Hamiltonians, where two potential energy surfaces cross. Similar to Dirac cones in graphene~\cite{geim2007}, these intersections are protected by symmetries of the Hamiltonian which guarantee that any loop in parameter space around a conical intersection has a quantized Berry phase~\cite{berry1984}. CIs play an important role in photochemistry \cite{domcke2011,yarkony2012}, as they mediate reactions such as photoisomerization and non-radiative relaxation, which are key steps in processes such as vision~\cite{polli2010} and photosynthesis \cite{olaso-gonzalez2006}. Therefore, detecting the presence and resolving the properties of CIs is important for computing reaction and branching rates in photochemical reactions \cite{Zimmerman1966, Bernardi1996}. Nevertheless, the study of such processes requires electronic structure methods capable of accurately modelling both the shape and the relative energies of the two intersecting potential energy surfaces, a requirement that poses challenges for the current available methods \cite{Gonzalez2012}. Given the need to develop novel methods for identifying and characterizing CIs, quantum computers present themselves as a highly promising option for this task.

Quantum computing has long been driven by the desire to simulate interacting physical systems, such as molecules, as a novel means of investigating their properties~\cite{feynman1982,aspuru-guzik2005}.
This is typically achieved by preparing eigenstates of molecular Hamiltonians in quantum devices which can natively store and process quantum states. 
This task would otherwise require an exponentially-scaling classical memory.
Recently, with the first noisy and intermediate-scale quantum devices (NISQ) \cite{preskill2018} being built, it became increasingly important to research tailored and robust algorithms that minimize the quantum device requirements~\cite{preskill2018}. 
Variational quantum algorithms (VQA), such as the variational quantum eigensolver~\cite{peruzzo2014,mcclean2016} (VQE) and its variations, caught the spotlight in this context, as they allow to prepare and measure quantum states with circuits of relatively low depth. 
The key feature of VQAs is the repeated execution of short parameterized quantum circuits on the quantum device, from which measurement results are sampled. 
These results are used to estimate a cost function, which is then minimized by varying the parameters defining the gates of the quantum circuit. 
Due to the noise introduced by sampling, a relatively large number of circuit runs and measurements are typically needed to estimate the cost function accurately. 
In chemistry, where VQEs are often proposed as a method to resolve ground state energies to high accuracy, the number of required samples to achieve such accuracy can become prohibitively large~\cite{wecker2015}.
Furthermore, the convergence of the cost function to an optimum is typically only suggested heuristically, and it is proven to be problematic in some cases that lack such heuristic structure~\cite{mcclean2018}. 
Therefore, it is compelling to suggest VQAs that can access quantities that are less reliant on the precision of both the optimization process and the measurement procedure.

A promising target for VQAs is the computation of the Berry phase $\Pi_\C$, which can be used to resolve the existence of CIs. 
More specifically, $\Pi_C$ is defined as the geometric phase acquired by an eigenstate of a parameterized Hamiltonian over a closed adiabatic path $\C$ in parameter space~\cite{berry1984}. 
Most importantly, it is known that in the presence of certain symmetries, the Berry phase will be quantized to values $0$ or $\pi$. 
This quantization is exactly what makes the Berry phase an attractive target for a VQAs, as it implies the final result of the computation need only be accurate to error $<\frac{\pi}{2}$. 
Quantum algorithms to compute Berry phases have been already proposed, both variational~\cite{tamiya2020,xiao2022} and Hamiltonian-evolution based~\cite{murta2020}. 
Moreover, the long-known effects of Berry phase on nuclear dynamics around a conical intersection \cite{longuet-higgins1958,mead1979a,ryabinkin2017} have been explored recently in analog quantum simulation experiments \cite{whitlow2023,valahu2023,wang2023}.
Nevertheless, previous proposals did not attempt to detect CIs in realistic quantum chemistry problems with an efficient algorithm that can be run on NISQ devices.

In this paper, we propose a hybrid quantum algorithm to compute the quantized Berry phase for ground states of a family of parameterized real Hamiltonians. 
We focus on the specific application to molecular Hamiltonians, where we can identify a conical intersection by measuring the Berry phase along a loop in atomic coordinates space. We first review the definition of CIs and Berry phases in Sec.~\ref{sec:background}.
Then, we present all the ingredients of the proposed algorithm in Sec.~\ref{sec:method}, which is similar to a VQA in spirit, but it does not require full optimization; rather, the variational parameters are updated by a single Newton-Raphson step for each molecular geometry along a discretization of the loop. 
In Sec.~\ref{sec:error-bounding}, we prove a convergence guarantee for the algorithm under certain assumptions on the ansatz, by providing sufficient condition bounds on the total number of steps and the acceptable sampling noise.
Finally, we adapt our algorithm to a specific ansatz in Sec.~\ref{sec:OOPQC-ansatz} and we benchmark it on a model of the formaldimine molecule \ce{H2C=NH} in Sec.~\ref{sec:numerics}. 
Section~\ref{sec:conclusion} presents our conclusions, a discussion of potential application cases for our algorithm and an outlook on possible enhancements.

The core code developed for the numerical benchmarks, which provides a flexible implementation of an orbital-optimized variational quantum ansatz, is made available in a GitHub repository at \url{https://github.com/emieeel/auto_oo} \cite{auto_oo}.

\section{Background}
\label{sec:background}

\subsection{Conical intersections}

Let us consider a molecular electronic structure Hamiltonian $H(\R)$ parameterized by the nuclear geometry $\R$ in some configuration space $\mathcal{R}$.
A conical intersection is a point $\R^\times \in \mathcal{R}$ where two potential energy surfaces become degenerate, leading to non-perturbatively large non-adiabatic couplings, and thus a breakdown of the Born-Oppenheimer approximation~\cite{teller1937,herzberg1963}.
Conical intersections extend to a manifold of dimension $\dim[\mathcal{R}]-2$, and lead to the two potential energy surfaces taking the form of cones in the remaining two directions $\hat{x}$, $\hat{z}$. 
These two potential energy surfaces can be described as eigenstates of the effective Hamiltonian 
\begin{equation}\label{eq:effective-hamiltonian}
    H^\text{eff}(\R)=h_x [\R - \R^\times]_x \sigma_x + h_z [\R - \R^\times]_z \sigma_z.
\end{equation}
The Pauli terms $\sigma_x$ and $\sigma_z$ form a complete basis for two-dimensional real symmetric matrices.
Thus, a single conical intersection cannot be lifted by any real-valued and continuous perturbation of $H^\text{eff}(\R)$; such a perturbation would only shift the value of $\R^\times$.
Moreover, the presence of such an effective Hamiltonian implies that in any direction other than $R_x$ or $R_z$ the energies must be degenerate.

Simulations involving conical intersections are challenging due to the degeneracy of the two states involved, as the character of both states needs to be considered.
Active space methods are often used for organic molecules, which involve selecting the chemical bonds that are formed or broken in the reaction pathway as well as the most significant spectator or correlating orbitals \cite{helgaker2000}. 
The situation is more complicated when transition metals are involved because the close energetic spacing of d-orbitals typically requires including all five d-orbitals of such a metal into the active space. 
An additional complication arises if the two crossing states correspond to different atomic configurations: a situation that is not uncommon for the early or late d- (or f-) metals, for which configurations with a different d- (f-) population are energetically close. 
In such cases, one may need to either work with non-orthogonal orbitals \cite{broer2003} or add an additional d-shell to the active space \cite{veryazov2011} to qualitatively describe the nature of both states.
For cases of practical interest in which one wants to characterize and simulate the internal conversion processes in a complex photo-excited system, the presence of transition metals may easily lead to large active space requirements. 
These can not be met by classical algorithms and would be highly challenging for quantum algorithms as well. 

One way to reduce the complexity of the problem is to first focus on the presence or absence of conical intersections that connect the ground and excited states.
The measurement of the Berry phase in chemical systems allows this: without explicitly computing the excited state surface and non-adiabatic couplings, it should be possible to detect whether a loop in the nuclear coordinate space encloses a conical intersection or not. 
In this manner, one may alleviate the requirements for the active space selection and orbital optimization and quickly establish the region in the potential energy surface that contains an intersection with another surface and needs to be scrutinized further~\cite{Yarkony1996}. 
Information about the location of conical intersections is of interest also for ground-state dynamics; the CIs and the Berry phase they induce influence the propagation of nuclear wave packets on the adiabatic ground state surface and thereby affect the branching rates and efficiency of reactions or isomerizations~\cite{aldenmead1980}.
For both types of applications, precise study of dynamics on ground state surfaces as well as characterizing the efficiency of radiationless decay, it is of interest to explore the possibilities offered by quantum algorithms.

\subsection{Berry phases in real Hamiltonians}

The Berry phase $\Pi_{\C}$ is the geometric phase acquired by some eigenstate $\ket{\eigstate(\R)}$ of a system with parameterized Hamiltonian $H(\R)$ as it is adiabatically transported around a closed loop in parameter space $\C \subset \mathcal{R}$.
$\Pi_{\C}$ can be defined as the closed line integral of the Berry connection along the loop $\C$
\begin{equation} \label{eq:berry-phase}
\Pi_{\C} = -i\oint_{\C}d\R\cdot\bra{\eigstate(\R)}\nabla_{\R}\ket{\eigstate(\R)}.
\end{equation}
The integrand must be imaginary, as $\nabla_{\R}\braket{\eigstate(\R)}{\eigstate(\R)} = 2\Re[\bra{\eigstate(\R)}\nabla_{\R}\ket{\eigstate(\R)}] = 0$, thus $\Pi_{\C}$ is real.
In this work, we assume $\ket{\eigstate(\R)}$ is the ground state of $H(\R)$.

We need to parameterize the loop $\C=\{\R(t),t\in[0,1]\}$ in order to evaluate the integral. 
Moreover, it is possible to multiply the ground state $|\eigstate(\R)\rangle$ by a $t$-dependent phase, resulting in a $U(1)$-gauge a transformation which leaves all physical quantities invariant. 
We take
\begin{equation}
\vert\gaugedstate(t)\rangle=e^{i\Theta(t)}\vert\eigstate(\R(t))\rangle,\label{eq:chi_prime_def}
\end{equation}
which allows us to rewrite the Berry phase as
\begin{equation}
\Pi_{\C}=-i\int_0^1 d t\langle\gaugedstate(t)\vert\partial_t\vert\gaugedstate(t)\rangle + \int_0^1d t\,\partial_t\Theta(t).
\end{equation}
If there is a representation for which each Hamiltonian $H(\R)$ is real,
it is possible to choose eigenstates that have all real components.
In this case, we can choose $\Theta(t)$ such that $|\gaugedstate(t)\rangle$ has real expansion, which implies the first integrand is real; as $\partial_t \langle\gaugedstate(t)|\gaugedstate(t)\rangle = 0$ this must also to be imaginary, therefore $\langle\gaugedstate(t)|\partial_t|\gaugedstate(t)\rangle=0$.
Under this choice, we can evaluate the Berry phase as a boundary term
\begin{equation} \label{eq:phi-boundary-term}
\Pi_{\C}=\int_0^1 d t\,\partial_t\Theta(t)=\Theta(1)-\Theta(0)=\mathrm{arg}\Big[\langle\gaugedstate(0)\vert\gaugedstate(1)\rangle\Big],
\end{equation}
where the last equality is obtained using the definition in Eq.~\eqref{eq:chi_prime_def}.
Furthermore, $\vert\gaugedstate(1)\rangle$ and $\vert\gaugedstate(0)\rangle$ are real by construction, which implies $\Pi_{\C}$ can only take two values (modulo $2\pi$): $0$ or $\pi$.

The quantization of $\Pi_{\C}$ implies that it is invariant for topological deformations of $\C$. 
If $\C$ can be contracted to a point, then $\Pi_\C=0$.
A non-trivial $\Pi_{\C}=\pi$ can only occur when $\C$ encircles a degeneracy.
One can check with the effective Hamiltonian Eq.~\eqref{eq:effective-hamiltonian} that any loop encircling $\R^\times$ has a Berry phase of $\pi$.
This extends by continuity to any region of $\mathcal{R}$ around the CI, as long as $\C$ does not enclose a second degeneracy point.
Thus, we have a one-to-one correspondence between CIs and the nontrivial Berry phase.

\subsection{Measuring Berry phase with a variational wavefunction}

There have been various proposals in the literature for computing Berry phases using a gate-based quantum device~\cite{murta2020,tamiya2020}.
In this work, we propose to use a variational algorithm to track $\vert\gaugedstate(t)\rangle$ when parallel-transported around the loop $\C$, in the spirit of the variational adiabatic method described in Ref.~\cite{wecker2015,harwood2022}.
We approximate 
\begin{equation}\label{eq:ansatz-state}
    \ket{\gaugedstate(t)} \approx \ket{\psi(\thetaex_t)}:= U(\thetaex_t)\ket{\psi_{0}},
\end{equation}
where $\ket{\psi(\bftheta)}$ is a variational ansatz state, and $\thetaex_t$ continuously tracks a local minimum [$\nabla_\bftheta E(t, \thetaex_t) = 0$] of the variational energy
\begin{equation}\label{eq:var_energy}
    E(t, \bftheta)=\langle\psi(\bftheta)|H(\R(t))|\psi(\bftheta)\rangle.
\end{equation}
The angle $\thetaex_t$ is well-defined as long as the Hessian $\nabla^2_{\bftheta} E(t, \bftheta)$ remains positive definite in a neighbourhood of $\thetaex_t$ for all $t$, ensuring the $\thetaex_t$ is continuous in $t$ and non-degenerate.
Although our treatment naturally extends to any variational ansatz that continuously parametrizes normalized states $\ket{\psi(\bftheta)}$ (including classical ansätze like e.g.~matrix-product states), we assume the operator $U(\bftheta)$ is implemented by a parameterized quantum circuit (PQC) acting on an initial state $\ket{\psi_0}$; this implies that information about the state needs to be extracted from a quantum device through sampling.

\section{Methods}
\label{sec:method}

In this section, we detail all the ingredients needed to implement our hybrid algorithm to resolve quantized Berry phases with a variational quantum ansatz. 
Initially, in Sec.~\ref{sec:real-ansatz}, we discuss how selecting an ansatz that preserves the Hamiltonian's symmetries establishes a natural gauge, leading to the reduction of the Berry phase integral to the boundary term Eq.~\eqref{eq:phi-boundary-term}.
In Sec.~\ref{sec:NR-update}, we introduce our parameter update approach, which employs single Newton-Raphson steps to trace the variational state along a discretization of the loop $\C$.
In Sec.~\ref{sec:regularization} we explain how to employ a basic regularisation technique to handle the potential non-convexity of the cost function and in \ref{sec:overlap} we explain how to measure the final overlap using an ancilla-free Hadamard test. 
Finally, in Sec.~\ref{sec:overview-agorithm}, we provide a full overview of the algorithm.

\subsection{Fixing the gauge with a real ansatz}
\label{sec:real-ansatz}

As discussed in Sec.~\ref{sec:background}, the quantization of the Berry phase is granted by the symmetries of the Hamiltonian family $H(\R)$, which ensure the existence of a basis for which each $H(\R)$ has a real representation. When it comes to electronic structure Hamiltonians, it is always possible to find a real representation for time-reversal symmetric Hamiltonians with integer total spin~\cite{mead1979}. Moreover, real noninteger-spin Hamiltonians are also found throughout nonrelativistic quantum chemistry.

As our variational ansatz state [in Eq.~\eqref{eq:ansatz-state}] is defined by a family of unitary operators, it inherits a natural gauge from $U(\bftheta)$.
In particular, if $U(\bftheta)$ is written as a product of real rotations in the basis in which $H(\R)$ is real, then we force $\ket{\psi(\bftheta)}$ to have real components as well, which fixes a global $U(1)$ phase. This can be obtained by constructing the PQC with a sequence of parameterized unitaries such as
\begin{equation}
    U_j(\theta_j) = e^{A_j \theta_j}
\end{equation}
generated by antisymmetric operators $A_j$ that are real in the chosen representation.
(We choose dimensional units such that $\lVert A_j \rVert = 1$ without loss of generality, see Appendix~\ref{app:parameter-change}).
Examples from electronic structure include real fermionic (de-)excitations, such as unitary singles $(A_{pq} = \hat{a}^{\dag}_p\hat{a}_q-\hat{a}^{\dag}_q\hat{a}_p)$ and doubles $(A_{pqrs} = \hat{a}^{\dag}_p\hat{a}_q \hat{a}^{\dag}_r\hat{a}_s - \hat{a}^{\dag}_q\hat{a}_p \hat{a}^{\dag}_s\hat{a}_r)$.
Many PQC ansätze commonly proposed for quantum chemistry, such as unitary coupled cluster (UCC) \cite{bartlett1989,peruzzo2014,romero2018} and quantum-number preserving gate fabrics (NPF) \cite{anselmetti2021}, are composed from these elementary rotations. 
Formally, our ansatz state can then be defined as
\begin{equation} \label{eq:real-unitary-ansatz}
    \ket{\psi(\bftheta)} = \prod_{j=1}^{n_p} U_{n_p - j}(\theta_{n_p - j}) 
    \ket{\psi_0},
\end{equation}
where $U_k$ are the aforementioned parameterized rotations applied in circuit-composition order.

In this case, the Berry phase can be estimated by $\Pi_{\mathcal{C}}=\arg\Big[\braket{\psi(\thetaex_{t=0})}{\psi(\thetaex_{t=1})}\Big]$, which corresponds to the boundary term in Eq.~\eqref{eq:phi-boundary-term}. This implies that, in the case of a non-trivial Berry phase, the path traced by $\bftheta^*_t$ will not close up on itself (i.e. $\bftheta^*_1 \neq \bftheta^*_0$), highlighting an important difference between the ansatz parameters $\bftheta$, which fix the gauge of $\ket{\psi(\bftheta)}$, and the Hamiltonian parameters $\R$, which define a ground state $\ket{\eigstate(\R)}$ up to a $U(1)$ gauge freedom. However, the change in optimal parameters is insufficient to prove the existence of a nontrivial Berry phase; we must both successfully track the minimum $\thetaex_t$ as $t$ traces from $0$ to $1$, and estimate the final overlap $\langle\psi(\thetaex_0)|\psi(\thetaex_1)\rangle$. While the argument (sign) of the overlap will yield the Berry phase, its absolute value is a proxy of success as it certifies the initial and final states are physically equivalent.

\subsection{Avoiding full optimization via Newton-Raphson steps}
\label{sec:NR-update}

As mentioned above, the Berry phase $\Pi_{\mathcal{C}}$ is a discrete quantity, and therefore we only need to estimate it to accuracy $<\frac{\pi}{2}$. 
In Appendix~\ref{app:parameter-change}, we show that this implies that we can accept an error on the estimate $\thetaest_1$ of the final optimum $\thetaex_1$ bounded in one-norm by $\lVert\thetaest_1 - \thetaex_1 \rVert_1 < 1$. Thus, we are not required to exactly track $\ket{\psi(\bftheta_t^*)}$ and as a result, the variational energy Eq.~\eqref{eq:var_energy} does not need to be fully re-optimized at every time-step $t$. Instead, it suffices to keep the estimate $\thetaest_t$ of the optimal parameters within the basis of convergence of the true minimum $\thetaex_t$.

To achieve this, we still need an initial optimum as an input, which is obtained by running one full optimization. If possible, the initial point is selected such that optimization is simplest. Then, we propose to use a single step of the Newton-Raphson algorithm  at points $t\in\{\Delta t, 2\,\Delta t,\ldots, 1-\Delta t, 1\}$.

The Newton-Raphson (NR) algorithm determines the update of the estimate of the minimum $\thetaest$ through the gradient $\mathcal{G}(t, \bftheta) := \nabla_\bftheta E(t, \bftheta)$ and Hessian $\mathcal{H}(t, \bftheta) := \nabla^2_\bftheta E(t, \bftheta)$ of the variational energy Eq.~\eqref{eq:var_energy}.
The derivatives can be computed using either finite-difference methods or parameter-shift rules \cite{schuld2019}; either method requires sampling the variational energy $E(t,\bftheta)$ at a number of different parameter points $\bftheta$.
Given the estimate $\thetaest_t$ of the optimum at point $t$ as an initial guess, the NR step with cost $E(t+\Delta t, \bftheta)$ prescribes the update $\thetaest_{t+\Delta t} = \tilde{\bftheta}_{t} + d\bftheta^\text{NR}_{t, \Delta t}$, with
\begin{equation} \label{eq:nr-step-definition}
    d\bftheta^\text{NR}_{t, \Delta t} = -\mathcal{H}^{-1}(t+\Delta t, \thetaest_t) \boldsymbol{\mathcal{G}}(t+\Delta t, \thetaest_t).
\end{equation} 

The Newton-Raphson method is well-known to have a finite-sized basin of quadratic convergence, as long as the cost function is strongly convex at the optimum,
\begin{equation} \label{eq:strong-convexity-at-minimum}
        m(\thetaex) := 
        \min_\mathbf{v} \frac{\mathbf{v}^T \mathcal{H}(\thetaex) \mathbf{v}}{\mathbf{v}^T \mathbf{v}} 
        > 0.
\end{equation}
In other words, the lowest eigenvalue $m$ of the Hessian of the cost function (which we call convexity) at the optimum $\bftheta^*$ needs to be positive.
Details on the convergence properties of NR are given in App.~\ref{app:proof-NR-error-bound}.
In section \ref{sec:error-bounding}, we show that this quadratic convergence is fast enough to keep track of the minimum with only a single step for each $t$-point.

\subsection{Regularization and backtracking}
\label{sec:regularization}

To ensure that we successfully track the minimum of the cost function $\thetaex_t$, the estimates $\thetaest_t$ need to remain in the strongly convex region of optimization space. 
The existence of such strongly convex region is not always guaranteed since it depends on the ansatz. 
A common cause of failure of this requirement is exemplified for ansätze with (local) over-parametrization of the state manifold. 
In fact, if the ansatz has redundant parameters the Hessian of the cost function Eq.~\eqref{eq:var_energy} will always be singular ($m=0$). Then, arbitrarily small perturbations of the cost function can then cause $m<0$.
When this occurs, the inversion of the Hessian needed for the Newton-Raphson step is ill-defined. 

The most direct approach to solve this issue is to select an ansatz with no degeneracies, facilitating a strongly convex cost function at its minima. 
Nevertheless, this is only possible for  very simple problems, and even in this case, quasi-degeneracies can make the convergence region extremely small.
Since this is a well-known problem, many alternative solutions have been proposed in the literature. 
In this subsection, we will explore and implement two of them -- back-tracking and regularization.

\emph{Back-tracking ---}
Small positive eigenvalues of the Hessian can cause the standard Newton-Raphson step to overshoot along the relative parameter eigenmodes.
This effectively reduces the size of the neighborhood of the minimum $\theta^*$ in which quadratic convergence is granted (see Appendix~\ref{app:proof-NR-error-bound}).
Since for positive convexity the direction provided by the Newton-Raphson step is guaranteed to be a descent direction, we can mitigate this overshoot by implementing line-search of the minimum on the segment defined by the NR step.
In the common variant of back-tracking, the Newton step is iteratively damped by a constant $\beta \in (0, 1)$.
At each iteration, the cost function in the new point is measured, until the cost function is reduced enough (the detailed condition is given in Algorithm~\ref{alg:NR-regularized}).
While some additional evaluations of the cost function are needed, the (more expensive) gradient and Hessian are only calculated once.
Due to the repeated evaluation, one needs to consider extending line-search methods to cost functions evaluated with sampling noise.

\emph{Regularization ---}
For realistic ansätze and Hamiltonians, it is difficult to avoid (quasi-) redundancies in some regions of parameter space.
In this case, the cost function might not be strongly convex around the minimum, or the convexity might be too small to ensure a sufficiently-large convergence region. 
Furthermore, even for an ideally-convex cost function, noisy evaluation on a quantum device might result in a distorted Hessian with non-positive eigenvalues.
To mitigate this issue, we can use a regularization technique that penalizes the change in parameters along the quasi-redundant directions.
We propose to use augmentation of the Hessian to regularize the NR step, obtaining a so-called quasi-Newton optimizer. 
Hessian augmentation is a common practice in quantum chemistry methods that feature orbital optimization, such as self-consistent field methods \cite{helgaker2000}. 
If the smallest eigenvalue of the Hessian $\lambda_0$ is smaller than a positive threshold convexity $m_\text{thr}$, we construct the augmented Hessian as follows
\begin{equation}
    \mathcal{B} = \mathcal{H} + \nu \mathbb{1},
\end{equation}
where we add a constant $\nu > |\lambda_0|$.
The augmented Hessian is then positive, and we can realize the NR update as
\begin{equation}
    d\bftheta^{\rm NR} = - \beta \mathcal{B}^{-1} \boldsymbol{\mathcal{G}}.
\end{equation}
Here, $\beta$ is the damping constant from back-tracking line search and $\mathcal{G}$ is the gradient as in eq.~\eqref{eq:nr-step-definition}.
Regularization and back-tracking are typically used in tandem, as quadratic convergence is harder to guarantee when using regularization.
The choice of $\nu$ is non-trivial: we want it to be large enough to suppress parameter changes along quasi-redundant directions, but we need to avoid exaggerating the damping along relevant directions.
Common solutions include choosing $\nu = \rho |\lambda_0| + \mu$ with fixed positive constants $\rho$ and $\mu$, or using a trust-region method~\cite{jensen1984, helmich-paris2021} where the Newton step is constrained to lie within a ball of some radius $h$ (such that $||d\bftheta^{\rm NR}||_2 \leq h$).

The augmented Hessian method does not require further evaluations of the cost function, although the damping of the parameter updates might imply that more $t$-steps are needed to successfully resolve the Berry phase.

\begin{algorithm}
    \caption{NR step subroutine with regularization and backtracking}
    \label{alg:NR-regularized}
    
    \KwIn{%
    Initial estimated parameters $\thetaest_{\tau}$ \newline
    Estimated Hessian $\tilde{\mathcal{H}}$\newline
    Estimated gradient $\tilde{\boldsymbol{\mathcal{G}}}$\newline
    Cost function $E(\theta)$ given by eq.~\eqref{eq:var_energy} \newline
    Convexity requirement $m_\text{thr} > 0$ \newline
    Positive constants $\mu$, $\rho$, $\alpha$, $\beta$
    }
    
    \KwOut{Updated estimated parameters $\thetaest_{\tau} + d\bftheta$.}
    
    \vspace{1em}
    
    $\lambda_0 \gets $ lowest eigenvalue of $\tilde{\mathcal{H}}$\;
    
    \tcp{Regularization}
    \eIf{$\lambda_0 < m$}{
        $\mathcal{B} \gets \tilde{\mathcal{H}} + (\rho|\lambda_0| + \mu)\mathbb{1}$\;
        $d\bftheta \gets \mathcal{B}^{-1}\tilde{\boldsymbol{\mathcal{G}}}$
    }{
        $d\bftheta \gets \tilde{\mathcal{H}}^{-1}\tilde{\boldsymbol{\mathcal{G}}}$
    }
    
    \tcp{Backtracking}
    \While{
        $E(\thetaest_{\tau} + d\bftheta) > E(\thetaest_{\tau}) + \alpha (\boldsymbol{\tilde{\mathcal{G}}} \cdot d\bftheta)$}{
        $d\bftheta \gets \beta d\bftheta$
    }
    
    \Return $\thetaest_{\tau} + d\bftheta$
    
\end{algorithm}

\subsection{Measuring the final overlap}
\label{sec:overlap}

For a real ansatz, the overlap of the tracked states at $t=0$ and $t=1$ must be real and it can be rewritten as
\begin{equation} \label{eq:final-overlap}
    \braket{\psi(\thetaest_0)}{\psi(\thetaest_1)} = 
    \Re\left[\bra{0} U^\dag(\thetaest_0) U(\thetaest_1) \ket{0}\right].
\end{equation}
This quantity can readily be measured by a Hadamard test, implemented as
\begin{equation*}
     \Qcircuit @C=1.2em @R=0.6em {
         &&&&\mbox{X} \\
         \lstick{\ket{0}} & \gate{H} & \ctrl{1} & \ctrl{1} & \meter \\
         \lstick{\ket{\psi_0}} & {/}\qw & \gate{U^{\dag}({\thetaest}_0)} & \gate{U({\thetaest}_1)} & \qw \mbox{\qquad.}}
\end{equation*}
The required number of samples is small, as we only need to resolve whether the sign of the overlap is $+1$ (trivial $\Pi_\C=0$) or $-1$ (nontrivial $\Pi_\C=\pi$).

Implementing the circuit above requires to realize controlled-$U(\bftheta)$, which might increase significantly the depth of the compiled quantum circuit and make the implementation unfeasible on near-term hardware.
However, this requirement can be bypassed by using the control-free echo verification technique \cite{OBrien2021, Polla2023}, in place of the standard Hadamard test, to sample Eq.~\eqref{eq:final-overlap}.
This technique requires access to a reference state $\ket{\psi_\text{ref}}$ orthogonal to $\ket{\psi_0}$, which should acquire a known eigenphase $\varphi$ under the action of the PQC, $U(\bftheta)\ket{\psi_\text{ref}} = e^{i\varphi}\ket{\psi_\text{ref}}$.
As most of the PQC ansätze used for electronic structure states preserve the total number of electrons (including UCC and NPF), the fully unoccupied state $\ket{0...0}$ can be used as reference.
Control-free echo verification circuits only require implementing the non-controlled $U(\bftheta)$, and furthermore provide built-in error mitigation power.

\subsection{Overview of the algorithm}
\label{sec:overview-agorithm}

We are now ready to formalize the proposed algorithm for resolving Berry phases, Algorithm~\ref{alg:main}.
The formalization we present here will allow us to bound the number of steps and the sampling cost in the following Section~\ref{sec:error-bounding}.
Given a path $\C$ and a number of steps $1/\Delta t$, the algorithm attempts to calculate $\Pi_\mathcal{C}$ yielding either $\Pi_\mathcal{C} = 0$, $\Pi_\mathcal{C} = \pi$, or a FAIL state. Again, in Section~\ref{sec:error-bounding} we will bound the probability of the FAIL state occurring.
Additional features that extend the practicality of the algorithm and mitigate the failure cases are presented in Sec.~\ref{sec:OOPQC-ansatz} and later implemented in Sec.~\ref{sec:numerics}.

\begin{algorithm}[t]
    \caption{Resolve quantized Berry phase}
    \label{alg:main}
    
    \KwIn{%
    Family of Hamiltonians $H(\R)$, $\R\in\mathcal{R}$\newline
    Real ansatz PQC $\ket{\psi(\bftheta)}$\newline
    Set of initial optimal $\thetaex_0$ \newline
    Closed path $\C\in\mathcal{R}$, $\R(t): [0, 1] \mapsto \C$\newline
    Number of steps $N$ to discretize $\C$\newline
    Precision requirements $\sigma_\mathcal{G} > 0$ and $\sigma_\mathcal{H} > 0$\newline
    Convexity requirement $m_\text{thr} > 0$\newline
    Regularization $ \text{reg} \in \{\text{False}, \text{True}\}$ \newline
    Final fidelity requirement $F \in (0, 1)$
    }
    
    \KwOut{$\Pi_\C = 0$ or $\Pi_\C = \pi$ or \emph{FAIL}.}
    
    \vspace{1em}
    $\Delta t = 1/N$\;
    $\thetaest_0 = \thetaex_0$\;
    \For{$\tau \in \{0, \Delta t, 2\,\Delta t, ..., 1 - \,\Delta t\}$}{
        $E(t, \theta) \gets$ define cost function as in Eq.~\eqref{eq:var_energy}\;
        $\tilde{\boldsymbol{\mathcal{G}}}_j \gets$ sample the gradient to precision $\sigma_\mathcal{G}$\\
            $\qquad \left[\tilde{\boldsymbol{\mathcal{G}}}_j = \frac{\partial{E}}{\partial \theta_j}(\tau + \Delta t, \thetaest_\tau)\right]$\;
        $\tilde{\mathcal{H}}_{jk} \gets$ sample the Hessian to precision $\sigma_\mathcal{H}$ \\
            $\qquad\left[\mathcal{H}_{jk} = \frac{\partial{E}}{\partial \theta_j\partial \theta_k}(\tau + \Delta t, \thetaest_\tau)\right]$\; 
        \If{$\text{reg} = \text{False}$}{
            $\lambda_0 \gets$ lowest eigenvalue of $\tilde{\mathcal{H}}$\;
            \lIf{$\lambda_0 < m_\text{thr}$}{\Return \emph{FAIL} and exit}
            $d\bftheta^\text{NR} \gets -\tilde{\mathcal{H}}^{-1} \tilde{\boldsymbol{\mathcal{G}}}$\ (see Eq.~\eqref{eq:nr-step-definition});
            $\thetaest_{\tau+\Delta t} \gets \thetaest_{\tau} + d\bftheta^\text{NR}$\;
        }
        \If{$\text{reg} = \text{True}$}{
            $\thetaest_{\tau+\Delta t} \gets$ \textbf{Subroutine 1} ($\thetaest_\tau$, $\tilde{\boldsymbol{\mathcal{G}}}$, $\tilde{\mathcal{H}}$)\;}
    }
    $f \gets$ final overlap as in Eq.~\eqref{eq:final-overlap} to precision $F$\;
    \lIf{$f^2 < F$}{\Return \emph{FAIL} and exit}
    \Return $\Pi_\C = \arg\{f\}$
\end{algorithm}

If the algorithm fails, it can be re-run with a larger number of steps $N$ and thus a smaller step size $\Delta t$.
A smaller step size decreases additive NR error bound (the error per step scales as $\Delta t^2$, the total bound thus scales as $\Delta t$).

\section{Error analysis and bounding}
\label{sec:error-bounding}

In this section, we find analytic upper bounds on the cost of estimating a quantized Berry phase $\Pi_\mathcal{C}$ on a fixed curve $\mathcal{C}$ using Algorithm~\ref{alg:main}.
To simplify the treatment regularization and back-tracking are not considered: 
instead, we require each estimate $\thetaest_{j\,\Delta t}$ at the $j$-th step to be within the region of quadratic convergence of the cost function at the next t-step $E((j+1)\Delta t, \bftheta)$.
We prove that this translates to a guarantee of convergence of the algorithm, under three conditions:
\begin{enumerate}
    \item At the local minimum $\thetaex_t$, where $\psi(\thetaex_t)$ approximates the ground state, the cost function $E(t, \bftheta)$ is strongly convex [as described in Eq.~\eqref{eq:strong-convexity-at-minimum}];
    \item The number of discretization steps $N$ is sufficiently large;
    \item The sampling noise on each of the Hessian and gradient elements ($\sigma_\mathcal{H}$ and $\sigma_\mathcal{G}$ respectively) is sufficiently small.
\end{enumerate}
The first point entails a requirement on the cost function, defined by the family of Hamiltonians $H(\R)$ and the choice of ansatz $\ket{\psi(\bftheta)}$.
This requirement is not satisfied if the ansatz state is defined with redundant parameters.
We contend that, while strong convexity is a significant assumption, incorporating regularization (or one of the other techniques suggested in the outlook) can alleviate the necessity for such an assumption in practical applications.
Our proof provides upper bounds on $N$ and lower bounds on $\sigma_\mathcal{H}$ and $\sigma_\mathcal{G}$, which suffice to grant convergence.
However, these are not to be considered practical prescriptions, as we do not believe them to be optimal; rather they show which are the relevant factors playing a role in the convergence of the algorithm.
As the sampled gradient and Hessian are random variables, the guarantee of convergence for bounded error is to be understood in a probabilistic sense.

We first clarify natural assumptions and notation used in the calculation of the basin of convergence of Newton's method.
We require the cost function $E(t, \bftheta)$ to be twice-differentiable by $\bftheta$, for all $t$, in a region around the true minima $\thetaex_t$.
We require the Hessian to be Lipschitz continuous across this region,
\begin{equation} \label{eq:lipschitz-continuity}
    \lVert\mathcal{H}(t, \bftheta) - \mathcal{H}(t, \bftheta + d\bftheta) \rVert < L \lVert d\bftheta \rVert.
\end{equation} 
(Here, the Lipschitz constant $L$ can be considered a bound $\lVert \mathcal{T} \rVert \leq L$ on the norm of the tensor of third derivatives $T(t, \bftheta) = \nabla_\bftheta\nabla_\bftheta\nabla_\bftheta E(t, \bftheta)$.)
We also require that the gradient of the $t$-derivative $\dot{\boldsymbol{\mathcal{G}}} = \nabla_{\bftheta} \frac{dE}{d t}$ is bounded by $\dot{\mathcal{G}}_\text{max}$ in 2-norm.
These regularity conditions are satisfied for the PQC ansätze we consider in Sec.~\ref{sec:OOPQC-ansatz}, and in App.~\ref{app:norm-of-derivatives} we argue for bounds on $L$ and $\dot{\mathcal{G}}_\text{max}$.
The strong convexity assumption described in the previous paragraph entails a constant lower bound $m_\text{thr}$ on the smallest eigenvalue of the Hessian
$\mathcal{H}(t, \thetaex_t)$ at the minimizer $\thetaex_t$ for all $t$.

\subsection{Bounding the NR error}

We will first calculate a lower bound on $\Delta t$ that ensures the error $\delta\thetaest_t = \thetaest_t - \thetaex_t$ is bounded by a constant for all values of $t$. 
(as shown in Appendix~\ref{app:parameter-change}, a bound on $\lVert\delta\thetaest_{1}\rVert_1$ is sufficient to ensure $\Pi_\C$ can be accurately resolved.)
In this calculation, we will allow for an additive error $\sigma_{\boldsymbol{\theta}}$ on $\thetaest_t$ due to sampling noise; we will simultaneously calculate an upper bound on $\sigma_{\boldsymbol{\theta}}$.
We sketch the calculation here and defer details to App.~\ref{app:proof-NR-error-bound}.

Firstly, it can be shown (Theorem~\ref{thm:quadratic-convergence} in Appendix~\ref{app:proof-NR-error-bound}) that the Newton-Raphson step Eq.~\eqref{eq:nr-step-definition} with cost function $E(t+\Delta t, \bftheta)$ is guaranteed to converge quadratically \cite{nocedal2006} to the minimizer $\thetaex_{t+\Delta t}$ as long as the initial guess $\thetaest_t$ is within a ball centred in the minimizer of radius $\frac{m_\text{thr}}{4L}$. 
Quadratic convergence means that the distance of the updated guess from the minimizer will scale as the square of the distance of the initial guess from the minimizer,
\begin{equation}
    \lVert \delta\thetaest_{t+\Delta t} \rVert \leq \frac{L}{m_\text{thr}} \lVert \thetaest_t - \thetaex_{t+\Delta t}\rVert^2.
\end{equation}
The right-hand side of this equation can be bounded through the triangle inequality as
\begin{equation}
\lVert \thetaest_t - \thetaex_t\rVert\leq \|\delta\thetaest_{t}\|+\|\thetaex_t-\thetaex_{t+\Delta t}\|.
\end{equation}
We can bound the second term in this equation by taking the total $t$-derivative of the optimality condition $\boldsymbol{\mathcal{G}}(t, \thetaex_t) = 0$, yielding 
\begin{equation}
\|\thetaex_{t+\Delta t}-\thetaex_t\| \leq m_\text{thr}^{-1}\dot{\mathcal{G}}_\text{max} \Delta t.
\end{equation}
If at step $t$ we are within the radius given by Theorem~\ref{thm:quadratic-convergence}, the NR step will quadratically converge, suppressing also the error from the previous step, and yielding $\|\delta\thetaest_{t}\|\leq \frac{m_\text{thr}^2}{16L^2}$.

To account for sampling noise effect, we then consider a small additive error $\sigma_\bftheta$ to $\thetaest_t$.
Maximising this allowed sampling noise at each step (as we will see, this becomes the bottleneck in our method) then yields the two bounds
\begin{equation} \label{eq:additive-error-and-timestep}
    \sigma_\bftheta \leq \frac{\sqrt{2} - 1}{4} \frac{m_\text{thr}}{L}
    , \quad 
    \Delta t \leq \frac{m_\text{thr}^2}{8 L \dot{\mathcal{G}}_\text{max}}.
\end{equation}
When both bounds are satisfied, we are guaranteed single steps of Newton's method will maintain convergence around the path $\mathcal{C}$.

\subsection{Bounding the sampling noise}

We now translate the bound on $\sigma_{\bftheta}$ to bounds on the variance of estimates of each element of the gradient and Hessian ($\sigma_{\mathcal{G}}^2$ and $\sigma_{\mathcal{H}}^2$ respectively).
This proceeds by simple propagation of variance through Eq.~\eqref{eq:nr-step-definition}.
We find
\begin{align} \label{eq:variance-propagation}     
    \sigma_{\boldsymbol{\theta}}^2:=\Var[d\bftheta^\text{NR}_{t, \Delta t}]
    \leq \quad&
    \lVert \mathcal{H}^{-1} \rVert^2 \mathbb{E}\left[
        \lVert \delta \boldsymbol{\mathcal{G}} \rVert^2
    \right] +
    \\ \nonumber +&
    \lVert \mathcal{H}^{-1} \rVert^2 \mathbb{E}\left[
        \lVert \delta\mathcal{H} \rVert^2
    \right] \lVert d\bftheta_{t, \Delta t}^\text{NR} \rVert^2,
\end{align}
where $\delta \boldsymbol{\mathcal{G}}$ and $\delta\mathcal{H}$ are the random variables representing the errors on gradient and Hessian. 
(a more detailed calculation is given in App.~\ref{app:sampling-noise}.)
Assuming $\bftheta$ has $n_p$ elements,  each element of the gradient is i.i.d. with variance $\sigma_{\mathcal{G}}^2$, we get 
\begin{equation}
    \mathbb{E}\left[ \lVert \delta \boldsymbol{\mathcal{G}} \rVert^2 \right] = n_p \, \sigma^2_{\mathcal{G}}.
\end{equation}
As $\delta\mathcal{H}$ is a $n_p \times n_p$ real symmetric matrix, assuming its elements are i.i.d. with variance $\sigma_{\mathcal{H}}^2$, we can invoke Wigner's semicircle law \cite{wigner1955} to approximate its norm by $\sqrt{n_p} \,\sigma_{\mathcal{H}}$, thus
\begin{equation}
    \mathbb{E}\left[ \lVert \delta \mathcal{H} \rVert^2 \right] \approx n_p \, \sigma^2_{\mathcal{H}}.
\end{equation}
Combining these with Eq.~\eqref{eq:variance-propagation}, and requiring the resulting variance to be small compared to the square allowed additive error $\sigma_\bftheta^2$ we obtain the bound
\begin{equation}
    \sigma^2_{\mathcal{G}}
    + \, \sigma^2_{\mathcal{H}} \lVert d\bftheta^\text{NR}_{t, \Delta t} \rVert^2 
    \ll
    \frac{3 - 2\sqrt{2}}{16}\frac{m_\text{thr}^4}{n_p L^2}.
\end{equation}
We can then bound the norm of the NR update as $\lVert d\bftheta^\text{NR}_{t, \Delta t} \rVert \leq m_\text{thr}^{-1} \lVert \boldsymbol{\mathcal{G}} \rVert$.
Splitting the error budget in half we obtain.
\begin{align}
    \sigma^2_{\mathcal{G}}
    &\ll
    \frac{3 - 2\sqrt{2}}{32}\frac{m_\text{thr}^4}{n_p L^2}
    \\
    \sigma^2_{\mathcal{H}}
    &\ll
    \frac{3 - 2\sqrt{2}}{32}\frac{m_\text{thr}^6}{n_p L^2 \lVert \boldsymbol{\mathcal{G}} \rVert^2}.
    \label{eq:hessian-variance-bound}
\end{align}
Note that these bounds are not tight. 
For instance, by applying Cauchy-Schwartz inequality to bound $\lVert \mathcal{H}^{-1} \cdot \mathcal{G} \rVert \leq \lVert \mathcal{H}^{-1} \rVert \lVert \mathcal{G} \rVert$, we overlook the fact that the gradient will change more slowly along a lower-eigenvalue eigenmode of the Hessian.
We believe further work might allow to define tighter bounds.

\subsection{Scaling of the total cost}
To give an estimate on how many measurements we need to sample gradient and Hessian to sufficient precision, we need to recast the quantities in Eq.~\eqref{eq:hessian-variance-bound} (the dominant term of the sampling variance) in terms of parameters of the problem.
If we use an ansatz without redundancies (or if we can get rid of redundancies through e.g.~regularization), and assuming we approximate the ground state well enough, the convexity $m_\text{thr}$ will be larger than the ground state gap $\Delta$, as every parameterized rotation in the PQC ansatz will introduce a state orthogonal to the ground state.
The norm of the gradient and the Lipschitz constant can be bound proportionally to their max norm, as shown in Appendix~\ref{app:norm-of-derivatives}, thus $\lVert \boldsymbol{\mathcal{G}} \rVert \leq \sqrt{n_p} \lVert H \rVert$, $\lVert \dot{\boldsymbol{\mathcal{G}}} \rVert \leq \sqrt{n_p} \lVert \dot{H} \rVert$ and $L \leq n_p^{3/2} \lVert H \rVert$ (where $\lVert H \rVert$ is the spectral norm of the Hamiltonian).
The number of measurements to sample the Hessian to precision Eq.~\eqref{eq:hessian-variance-bound} are proportional to the inverse of the bound, with proportionality constant $M_\mathcal{H}$ indicating the number of shots required to sample a single element of the Hessian to unit variance (this depends on details such as the decomposition taken to measure the Hamiltonian, and the specifics of the derivative estimation method).
Multiplying this by the number of steps $\frac{1}{d t}$ [Eq.~\eqref{eq:additive-error-and-timestep}] gives us the total number of shots required for convergence
\begin{align}
    M_\text{tot}
    &=
    10^3 \frac{
        n_p L^3 \lVert \boldsymbol{\mathcal{G}} \rVert^2 \dot{\mathcal{G}}_\text{max}
    }{\Delta^8} M_{\mathcal{H}}
    \label{eq:nshots_bound}
    \\
    & <
    10^3 \frac{
        n_p^4 \lVert H \rVert^7 \lVert \dot{H} \rVert
    }{\Delta^8} M_{\mathcal{H}}.
\end{align}

\section{Adapting to an orbital-optimized PQC ansatz}
\label{sec:OOPQC-ansatz}

To achieve a good representation of the ground state character while minimizing depth and number of evaluations of quantum circuits, we employ a hybrid ansatz composed of classical orbital rotations and a parameterized quantum circuit (PQC) to represent correlations within an active space.
The concept of an orbital-optimized variational quantum eigensolver (OO-VQE) is explored in \cite{Yalouz2021, yalouz2022analytical}, as an extension of VQE conceptually similar to a complete active space self-consistent field (CASSCF) calculation.
In this section, we introduce the construction of the OO-PQC ansatz and discuss its specific use in our algorithm, where the orbitals need to continuously track a changing active space depending on the nuclear geometry.

\subsection{An OO-PQC ansatz with geometric continuity}

To represent the electronic structure state, we start by choosing an atomic orbital basis, i.e.~a discretization of space defined by a set of $N$ non-orthogonal atomic orbitals $\chi_\mu(\R, \boldsymbol{x})$ (functions of the electronic coordinate $\boldsymbol{x} \in \mathbb{R}^3$, where we make explicit the parametric dependence on the nuclear coordinates $\R$);
these orbitals define the overlap matrix 
\begin{align}\label{eq:overlap-matrix}
    S_{\mu\nu}(\R) 
    =& 
    (\chi_\mu(\R) | \chi_\nu(\R)) 
    \\ :=& 
    \int_{\mathbb{R}^3} \chi_\mu^*(\R, \boldsymbol{x}) \chi_\nu(\R, \boldsymbol{x}) \, d^3\boldsymbol{x}.
\end{align}
The atomic orbitals (AOs), along with the overlap matrix, depend on the geometry of the molecule specified by the nuclear coordinates $\R$.
(For the sake of simplicity, we limit to considering real AOs.)
From these, we could define a set of parameterized orthonormal molecular orbitals (MO)
\begin{equation}
    \phi_p(\R, C^\text{AO}) = \sum_\mu \chi_\mu(\R) C^\text{AO}_{\mu p},
\end{equation}
which would allow for the definition of a parameterized active space.
The downside of this parametrization is that, to ensure MO orthonormality, we need $C^\text{AO}$ to satisfy the constraint
\begin{equation}\label{eq:mo-orthonormality-constraint}
    C^{\text{AO}\dagger}  S(\R)  C^\text{AO} = \mathbb{1},
\end{equation}
which depends nontrivially from $\R$.
This implies that we cannot trivially use the same $C^{\text{AO}}$ for different geometries $\R$.

In order to address this problem, we have opted to use orthonormalized atomic orbitals (OAO) that are derived from the AOs through symmetric Löwdin orthogonalization \cite{lowdin1950non} as reference in the definition of parameterized MOs.
The OAOs are defined as $\phi_p^\text{OAO}(\R) = \sum_\mu \chi_\mu(\R) S^{-1/2}_{\mu p}(\R)$.
Building on these, we can define the MOs as 
\begin{equation} \label{eq:parameterized-mo}
    \phi_q(\R, C) = \sum_{\mu, p} \chi_\mu(\R) S^{-1/2}_{\mu p}(\R) C_{pq},
\end{equation}
where $C = S^{1/2}(\R) \cdot C^\text{AO}$.
The orthonormality constraint Eq.~\eqref{eq:mo-orthonormality-constraint} then reduces to requiring $C$ to be orthogonal, and it is independent on $\R$.
In summary, Eq.~\eqref{eq:parameterized-mo} defines a set of orthonormal molecular orbitals parameterized by $C$, well-defined and continuous for changing $\R$.

To start up our algorithm, the matrix $C^\text{AO}$ can be initialized by a Hartree-Fock (or any other molecular coefficient matrix, e.g.~coming from a small CASSCF calculation) at some initial geometry $\R(0)$. 
From this we recover $C = S^{1/2}(\R(0)) \cdot C^\text{AO}$, which is then treated as a variational parameter of the ansatz.
Using the parameterized MO Eq.~\eqref{eq:parameterized-mo}  we construct the electronic structure Hamiltonian $H(R, C)$ in the (parameterized) molecular basis.

Based on the initial Hartree-Fock orbital energies, we split the $N$ orbitals into a core set with $N_\text{O}$ doubly-occupied orbitals, an active set with $N_\text{A}$ orbitals, and a virtual set with $N_\text{V}$ empty orbitals.
Although the split of the orbital indices remains constant throughout the algorithm, the orbitals themselves continuously change through their dependence on $\R$ and $C$.
The correlations are treated only within an active space of $\eta_\text{A}$ electrons in $N_\text{A}$ orbitals. 
Tracing out the core and virtual orbitals yields the active-space Hamiltonian $H_\text{A}(\R, C)$.

The correlated active-space state $\ket{\psi(\bftheta)}$ is represented on a quantum device, using a PQC ansatz of the form Eq.~\eqref{eq:real-unitary-ansatz}.
The cost function then becomes
\begin{equation}\label{eq:oovqecost}
    E(\R, C,\bftheta) = \bra{\psi(\bftheta)} H_\text{A}(\R, C) \ket{\psi(\bftheta)},
\end{equation}
and it can be evaluated by sampling the 1- and 2-electron reduced density matrix (RDM) of the state \cite{bonet-monroig2020}.
(Other efficient sampling schemes, e.g.~based on double factorization \cite{vonburg2021,cohn2021}, can be used.)

\subsection{Measuring boundary terms with the OO-PQC ansatz}
When evaluating the final overlap [Eq.~\eqref{eq:final-overlap}] with an orbital-optimized ansatz, we have to consider that the states $\ket{\psi(\bftheta_0)}$ and $\ket{\psi(\bftheta_1)}$ are defined on different active space orbitals, determined by the MO matrices $C_0$ and $C_1$ respectively.
The transformation between the two sets of orbitals, $\boldsymbol{\phi}(\R, C_1) = \boldsymbol{\phi}(\R, C_0) \cdot  C_{0\to1}$, is represented by the orthogonal matrix
\begin{equation}
    C_{0\to1} = C_0^\dag C_1.
\end{equation}
If the algorithms successfully tracked the lowest-energy active space state of the system, the Hilbert spaces spanned by the active orbitals defined by $C_0$ and $C_1$ should match.
(The same is true for the core space and the virtual space.)
This implies the matrix $C_{0\to1}$ will have block structure, with $[C_{0\to1}]_{pq} \neq 0 $ only if $p, q$ are both in the same set of orbitals (core, active or virtual).
The orbital rotations within the core (virtual) subspace will not generate any phase on the state of the system, as all orbitals are doubly-occupied (doubly-unoccupied).
The orbital rotation within the active space can then be translated to a unitary transformation on the state by a Bogoliubov transformation
\begin{equation}
    G_{0\to1} = 
    \exp
    \Big\{ 
        \sum_{\{p,q\}\in \text{AS}}
        [\log(C_{0\to1})]_{pq} c^\dag_p c_q
    \Big\},
\end{equation}
with $c^\dag_p, c_p$ fermionic creation and annihilation operators on the $p$ orbital.
The final overlap
\begin{equation} \label{eq:final-overlap-with-G}
    \Re\left[\bra{0} U^\dag(\thetaest_0) G_{0\to1} U(\thetaest_1) \ket{0}\right] =: \omega_\C
\end{equation}
can then be sampled with a Hadamard test, given an quantum circuit implementing the (eventually controlled) operation $G_{0\to1}$.
Under Jordan-Wigner encoding, a quantum circuit for $G_{0\to1}$ can be implemented as a fabric of parameterized fermionic swap gates of depth $N_\text{A}$ following a QR decomposition of  the orbital rotation generator $[\log(C_{0\to1})]_{pq}$, also known as a givens rotation fabric \cite{arute2020}.
These gates preserve the zero-electrons reference state, allowing to employ the ancilla-free echo verification technique mentioned in section \ref{sec:overlap} to measure the final overlap.

\subsection{Newton-Raphson updates of the OO-PQC ansatz}
\label{sec:OOPQC-NR-step}

The proposed OO ansatz has two sets of parameters, $C$ and $\bftheta$.
As the MO matrix $C$ is subject to the constraint Eq.~\eqref{eq:mo-orthonormality-constraint}, its elements cannot be freely updated with NR.
Instead, for each NR update with initial MO matrix $C$, we reparametrize the MOs with a unitary transformation: $C\leftarrow C e^{-\kappa}$, where $\kappa$ is any antisymmetric matrix.
The derivatives of the energy with respect to any element $\kappa_{pq}$ can be evaluated analytically (see Appendix~\ref{app:analyticOO}).
Furthermore, under this parametrization it can be shown that $\kappa_{pq}$ where $p,q$ are both core indices or both virtual indices are \emph{redundant} \cite{helgaker2000} in the definition of the active space orbitals; these $N_\text{O}^2 + N_\text{V}^2$ parameters are set to zero without reducing the expressivity of the ansatz.
We call the unraveled set of remaining parameters $\bfkappa$.
To implement the Newton-Raphson step, the gradient and Hessian with respect to the combined set of parameters $(\bftheta, \bfkappa)$ is computed. In this manner the gradient splits into two components, and the Hessian into three components
\begin{align}
    \nabla_{(\bftheta,\bfkappa)} E 
    &= (\nabla_{\bftheta}E, \nabla_{\bfkappa} E)  \label{eq:composite-gradient}
    \\
    \nabla^2_{(\bftheta,\bfkappa)} E 
    &= 
    \begin{bmatrix}
        \nabla^2_{\bftheta}E 
        & \nabla_{\bfkappa}\nabla_{\bftheta} E\\
        (\nabla_{\bfkappa}\nabla_{\bftheta} E)^\intercal
        & \nabla^2_{\bfkappa} E 
    \end{bmatrix}.  \label{eq:composite-hessian}
\end{align}
The PQC parameter derivatives $\nabla_{\bftheta}E$ and $\nabla^2_{\bftheta}E$ can be evaluated through parameter-shift rule~\cite{schuld2019,huembeli2021} or finite difference, by sampling on the quantum device.
The derivatives with respect to the OO parameters $\nabla_{\bfkappa}E$ and $\nabla^2_{\bfkappa}E$ are linear functions of the 2-electron RDM, whose coefficients can be computed analytically \cite{helgaker2000}.
The remaining component, the mixed Hessian $\nabla_{\bfkappa}\nabla_{\bftheta} E$ can be similarly evaluated as a linear function of $\bftheta$-derivatives of the RDM; for this, we can use the same data sampled from the quantum device to evaluate $\nabla_{\bftheta}E$.
We detail the procedure of estimating these Hessian components in  App.~\ref{app:analyticOO}.
Thus, evaluating the derivatives with respect to the OO parameters does not require extra sampling on the quantum device.

\section{Numerical results}
\label{sec:numerics}

\begin{figure*}[t]
    \includegraphics[width=0.94\textwidth]{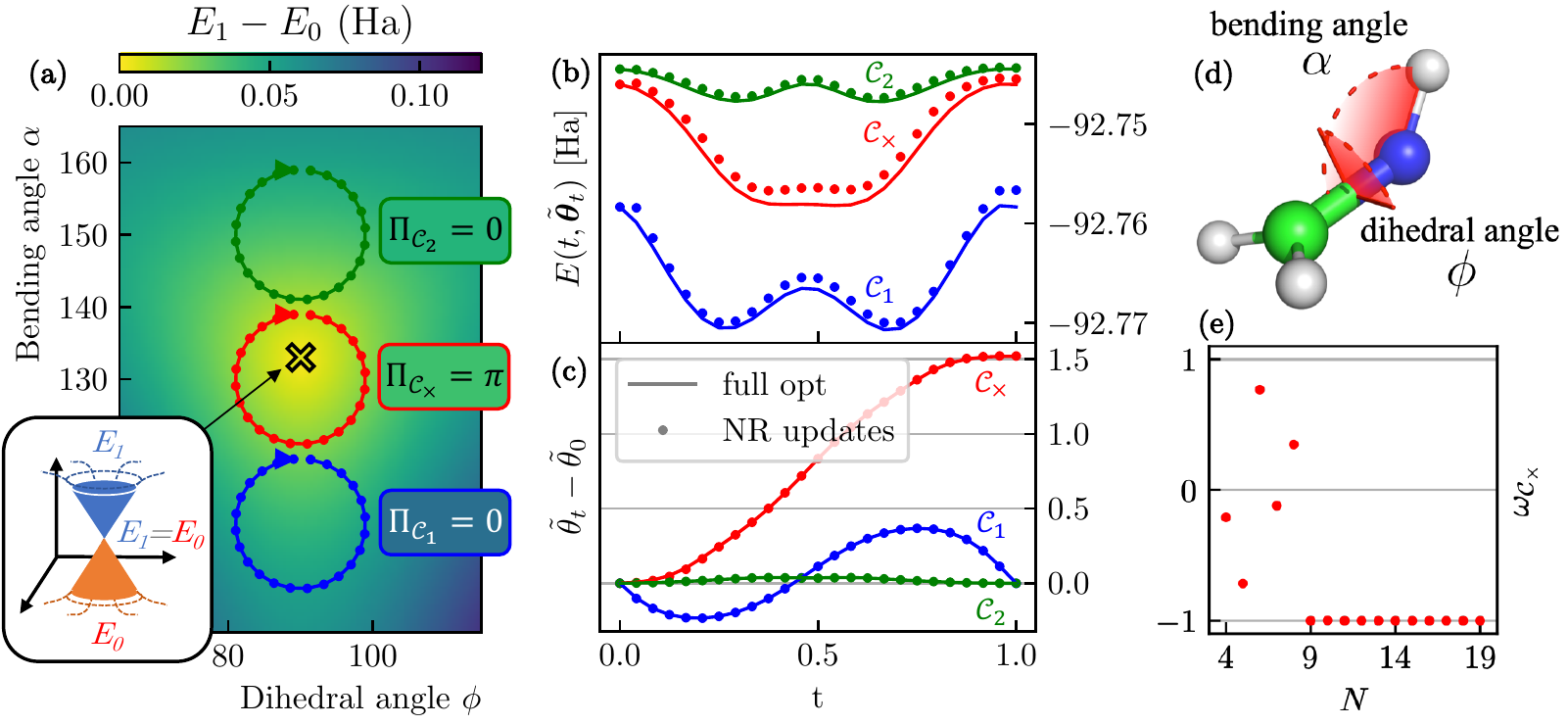}
    \caption{
        \textbf{(a)} Three loops in the nuclear configuration space of formaldimine; $\C_1$ (green), $\C_\times$ (red) and $\C_2$ (blue).
        $\C_\times$ encircles a CI, resulting in a non-trivial Berry phase.
        In this representation, the loops are discretized by $N=25$ points.
        The color plot indicates the energy gap at the full configuration interaction (FCI) level.
        \textbf{(b)} Energy and \textbf{(c)} change in PQC parameter around the three loops, with the same color coding as in (a) and the same $N=25$.
        These results refer to a (2,2) active space, and a minimal (STO-3G) basis set description of Formaldimine, with a OO-UCCD ansatz that has a single $\theta$ parametrizing the only double excitation.
        The continuous lines show the true optimum $\thetaex_t$ and the relative energy (obtained by full optimization), while the markers show the progress of the estimate $\thetaest_t$ from Algorithm~\ref{alg:main}, in absence of sampling noise.
        The Hessian stays positive throughout the path, no regularization is needed.
        \textbf{(e)} Final overlap computed by Algorithm~\ref{alg:main} for the red loop containing a CI, for a varying number of total discretization points $N$.
        For $N<9$, the Hessian is not always positive and regularization is needed to invert the Hessian, but no backtracking is used.
        \textbf{(d)} Schematic representation of Formaldimine, indicating the parameters used to define the nuclear geometries in this work.
     }
    \label{fig:CI-resolution}
\end{figure*}

In this section, we demonstrate the application of our method to a small model system: the formaldimine molecule \ce{H2C=NH}, an established model in the context of quantum algorithms for excited states in \cite{Yalouz2021, yalouz2022analytical, hirai2022}.
This molecule is known to have a conical intersection between the singlet ground state and first excited state potential energy surfaces \cite{bonacic-koutecky1985}.
This CI plays an important role in the photoisomerization process of formaldimine, which in turn can be considered a minimal model for the photoisomerization of the rhodopsin protonated Schiff-base (a key step in the visual cycle process \cite{birge1990, chahre1985}).
We consider geometries obtained from the equilibrium configuration by varying the direction of the \ce{N-H} bond, defined by the bending angle $\alpha$ and the dihedral angle $\phi$ (see Fig.~\ref{fig:CI-resolution}d).
Varying these angles defines the considered plane in nuclear configuration space $\mathcal{R}$.
First, we consider a minimal model of formaldimine (within the minimal basis and a small active space), on which we can test the properties of the algorithm \ref{alg:main}.
Then, we investigate the effects of sampling noise on these results.
Finally, we study a more complex model of the same molecule (with a larger basis set and active space), and show that we can achieve similar results by employing regularization and backtracking to deal with the degeneracies of the ansatz manifold.

\subsection{Numerical simulation details}
For our simulations, we use the PennyLane 
\cite{bergholm2022}
package with the PyTorch backend to construct the hybrid quantum-classical cost function of Eq.~\eqref{eq:oovqecost} supporting automatic differentiation (AD) with respect to the PQC parameters.
To achieve this, we implement the transformation of the one- and two-body AO basis integrals to the parameterized MO basis [Eq.~\eqref{eq:parameterized-mo}] with AD support.
The transformed integrals are projected onto the active space and contracted with the active space one- and two-electron RDM.
The RDM elements and their derivatives with respect to the PQC parameters are obtained using PennyLane and its built-in AD scheme based on the parameter shift rule.
In summary, gradients and Hessians of the cost function with respect to the PQC parameters are obtained using AD, the orbital gradient and Hessian components are estimated analytically (see Appendix~\ref{app:analyticOO}), and the off-diagonal block of the composite Hessian [Eq.~\eqref{eq:composite-hessian}] is retrieved by automatic differentiation of the analytical orbital gradient.
We use PySCF~\cite{sun2020} to generate the molecular integrals (i.e.~the full space one- and two-electron integrals and overlap matrices) in the atomic orbital basis.

The core code developed for this project is made available as a python package in the GitHub repository \cite{auto_oo}. 
This code provides a flexible implementation of the orbital-optimized PQC ansatz, which can find many applications in VQAs for chemistry.
A tutorial Jupyter notebook showcasing a calculation of Berry phase in the minimal model of Formaldimine is provided in the \url{examples} folder in the repository.

\subsection{Minimal model with an degeneracy-free ansatz}

We first demonstrate the application of our algorithm to a minimal model of formaldimine, for which we can approximate the ground state with a simple ansatz with no degeneracies.
The molecule is described in a minimal STO-3G basis-set, and we select an active space of $\eta_\text{A}=2$ electrons in $N_\text{A}=2$ spatial orbitals [i.e.~CAS(2,2)].
As the orbital optimization already allows (spin-adapted) single excitations within the active space, the only parameterized gate we can include in our PQC ansatz is the double-excitation $U(\theta) = e^{\theta (c_0^\dag c_1^\dag c_2 c_3 - c_2^\dag c_3^\dag c_0 c_1)}$; this corresponds to the unitary coupled-cluster doubles [UCC(S)D] ansatz, where the singles (S) are not explicitly included because they would be redundant with the orbital optimization.
This is enough to describe exactly any active space state compatible with the symmetries of the model, without over-parametrizing the ansatz state.

In Fig.~\ref{fig:CI-resolution} we demonstrate the application of our algorithm to this model. 
The minimal basis set is small enough that we can run a full configuration interaction (FCI) calculation to exactly resolve the ground and first excited state energies $E_0(\R)$ and $E_1(\R)$.
Observing the gap $E_1(\R) - E_0(\R)$ (portrayed in Fig.~\ref{fig:CI-resolution}a), we can determine the location of the conical intersection $\phi^\times = \ang{90}$, $\alpha^\times \approx \ang{132}$.
We then define three loops in the configuration space $\mathcal{R}$, one loop $\C_\times$ containing the CI and two ``trivial'' loops $\C_1$, $\C_2$.
These loops are centered around $\phi=\ang{90}$ and $\alpha=\ang{110}$ ($\C_1$), $\alpha=\ang{130}$ ($\C_\times$) and $\alpha=\ang{150}$ ($\C_2$), and all have a radius of $\ang{10}$.
Fig.~\ref{fig:CI-resolution}c shows the progress of the estimate $\tilde\theta_t$ (shifted by $\tilde\theta_0$) of the optimal PQC parameter $\theta^*_t$ throughout the $N=25$ single-NR-update t-steps.
Note that, while we only plot the single PQC parameter relative to the parameterized double excitation, the algorithm updates the MO coefficients $\tilde{C}_t$ as well.
We observe that $\thetaest_1 = \thetaest_0$ for the trivial loops $\C_1, \C_2$, while $\thetaest_1 \neq \thetaest_0$ for the loop $\C_\times$ containing the CI.
This is an indication of the effect of the Berry phase, but not the result of the algorithm yet; measuring the overlap Eq.~\eqref{eq:final-overlap-with-G} yields the correct Berry phase $\Pi_\C = \arg[\omega_\C]$.
The estimated energy $E(t, \thetaest_t)$ and the optimal $E(t, \thetaex_t)$ (obtained by full local optimization) are shown in Fig.~\ref{fig:CI-resolution}b. 
We can observe a small deviation from the optimal energy in the region where the character of the state changes faster (along the line $\phi=\phi^\times$, $\alpha<\alpha^\times$), but this does not disrupt the tracking of the minimum.
Finally, Fig.~\ref{fig:CI-resolution}e shows that a number of discretization points $N\geq9$ is needed to correctly resolve $\Pi_{\C_\times} = \pi$, through the evaluation of the overlap [Eq.~\eqref{eq:final-overlap-with-G}] $\omega_{\C_\times} = -1$.

\subsection{Sampling noise}

\begin{figure}[t]
	\centering
    \includegraphics[width=0.48\textwidth]{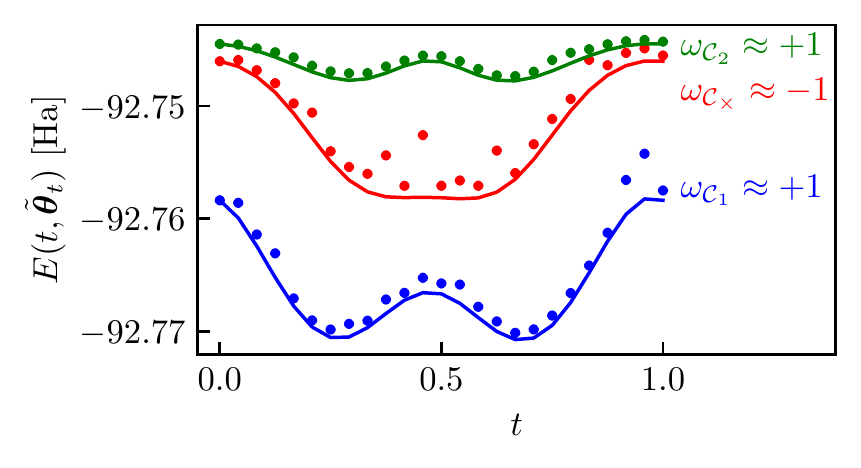}
    \caption{
        Energies throughout the loops in Fig.~\ref{fig:CI-resolution} (discretized by $N=25$ steps), in the presence of sampling noise. Each element of the composite gradient Eq.~\eqref{eq:composite-gradient} and Hessian Eq.~\eqref{eq:composite-hessian} are perturbed by random gaussian noise with variance $\sigma^2=\num{5e-6}$.
        In this instance the Hessian stays positive around the path, so no regularization is needed.
        The full lines, like in Fig.~\ref{fig:CI-resolution}b, indicate the true optimum $E(t, \thetaex_t)$ obtained by full optimization.
    }
    \label{fig:energies_noisy}
\end{figure}

In this section, we explore the robustness of our algorithm with respect to the sampling noise characteristic of VQAs.
We directly add a proxy of sampling noise $\eta$ to each element of the gradient and Hessian.
Each $\eta$ is an independent Gaussian random variable with variance $\sigma^2$; a different $\eta$ is added to each element of the gradient of \eqref{eq:composite-gradient} and Hessian of Eq.~\eqref{eq:composite-hessian} to get the noisy estimates $\tilde{\mathcal{G}}$ and $\tilde{\mathcal{H}}$.
In this way, one avoids defining a specific sampling strategy, keeping our results general. 

In Fig.~\ref{fig:energies_noisy} we show the energy profile of the three loops whose geometry is represented in Fig.~\ref{fig:CI-resolution}a, for one such random realisation of the sampling noise.
(The plotted energy expectation is evaluated exactly, noise is only added to the gradient and Hessian used in the NR updates.)
These three loops yield the same Berry phase results as the noiseless case.

\begin{figure}[t]
	\centering
    \includegraphics[width=0.48\textwidth]{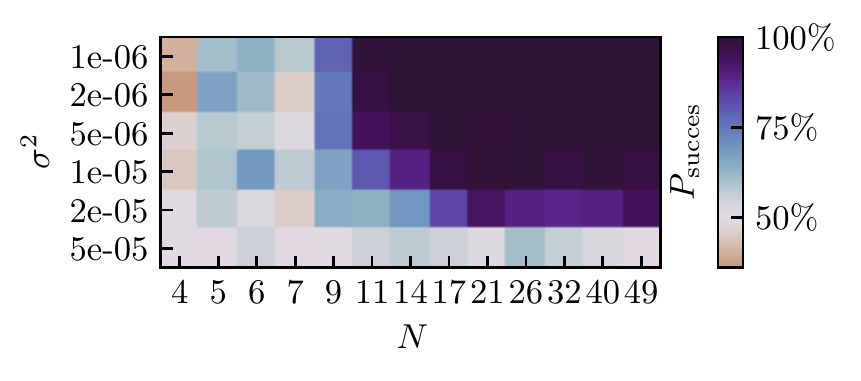}
    \caption{
        Success probability $P_{\rm succes}$ of the algorithm as a function of the number of discretization points $N$ and the sampling noise variance $\sigma^2$ for the loop $\C_\times$ containing a CI.
        The details of the model and geometry of the loop $\C_\times$ match Fig.~\ref{fig:CI-resolution} (red loop).
        Success is defined by resolving a final overlap $\braket{\psi(\thetaest_0)}{\psi(\thetaest_1)} < 0$, which returns $\Pi_{\C_\times}=\pi$. 
        The success probability is computed over 100 simulated runs. 
        A $P_\text{success} \approx 50\%$ indicates the algorithm returns random outcomes.
        Regularization by Hessian augmentation is enabled in these calculations, while no back-tracking is used.
    }
    \label{fig:success-noisy}
\end{figure}

The probability $P_\text{success}$ of Algorithm~\ref{alg:main} correctly resolving the Berry phase $\Pi_{\C_\times}$ on the nontrivial loop $\C_\times$ is reported in Fig.~\ref{fig:success-noisy}, as a function of the number of discretization steps $N$ and of the variance of the added noise on each sampled quantity $\sigma^2$.
The expected final overlap Eq.~\eqref{eq:final-overlap-with-G} is $-1$ for this case, as the loop contains a CI.
For each value of $N$ and $\sigma^2$, we simulate 100 noisy runs of the algorithm and we declare as successful the ones that yield a negative final overlap (implying $\Pi_{C_\times} = \pi$). Finally we average these outcomes to retrieve the succes probabilities $P_{\rm succes}$.

From these simulations we conclude that sampling noise reduces the accuracy of tracking the ground state and thus increases the probability of obtaining inaccurate energies. 
Nevertheless, for a moderate amount of sampling noise our algorithm still resolves the Berry phase correctly. 
We observe that an error on each gradient and Hessian element with variance of $\sigma^2 = 10^{-5}$ (or smaller) does not compromise the resolution of the Berry phase, as long as the number of discretization points is sufficiently large ($N>10$, very close to the noiseless case portrayed in Fig.~\ref{fig:CI-resolution}e).
On the other hand, a large enough sampling error ($\sigma^2 \geq \num{5e-5}$) produces essentially random results ($P_\text{success} \approx 50\%$).

The number of shots required to achieve a given accuracy on gradient and Hessian depends on the chosen measurement optimization scheme and the technique used to estimate derivatives.
We note that the energy can be expressed as a linear function of one- and two-electron RDMs,
\begin{equation}\label{eq:cost_rdm}
    E(t, \bftheta) = \sum_\xi D(\bftheta)_\xi h(t)_\xi,
\end{equation}
where $\xi = {pq, pqrs \in [N_\text{A}]}$ are the one- and two-body active-space orbital indices, $D(\bftheta)_\xi$ are the RDM elements and $h(t)_\xi$ are the one- and two-electron integrals.
The RDM elements can be grouped and measured together with optimized shot allocation \cite{bonet-monroig2020,huggins2021,zhao2021,choi2022,gresch2023}, with the potential for further optimization based on basis selection \cite{koridon2021} or on a more advanced factorization of the cost function \cite{cohn2021,hohenstein2022}.
Derivatives are also linear in the RDMs evaluated at different parameter values (see App.~\ref{app:analyticOO}), and they can be reconstructed either through finite-difference methods or using parameter shift rules \cite{schuld2019,wierichs2022generalparameter}.

To provide an upper bound on the number of shots necessary to achieve the required accuracy in our numerical benchmark, we consider an un-optimized measurement scheme which samples each element of the RDMs separately.
In this scheme, optimal resource allocation requires each $D(\bftheta)_\xi$ to be sampled with a number of shots proportional to $|h(t)_\xi|$ \cite{Rubin2018}.
This allows to estimate the cost function to accuracy $\sigma$ with a number of shots upper bounded by $N_\text{shots} = \sigma^{-2}||\vec{h}(t)||_1^2$, with $||\vec{h}(t)||_1 = \sum_\xi|h(t)_\xi|$ the total one-norm of the electronic integrals.
The Gradient and Hessian with respect to a single parameter are estimated through the parameter shift rule only requiring 3 parameter points.
The same RDM samples can be used to estimate all elements of the mixed gradient and Hessian, and we assume that the one-norm of the derivatives of $h$ with respect to the orbital rotations are of the same magnitude as $||\vec{h(t)}||_1$.
In our simulation benchmark (on the path $\C_\times$), the one-norm of the integral takes values up to $\max_t ||\vec{h}(t)||_1 = 1.5376$.
Thus, we estimate an upper bound on the total number of shots of the order of $\approx 3 \times 2.4 \sigma^{-2}$.

\subsection{Larger basis and active space} 
\label{sec:larger-system}

To test convergence for a more realistic case where the cost function is not always strongly convex at its minima, we simulate the algorithm on a more challenging model of formaldimine.
The model is constructed employing the cc-pVDZ basis set (43 atomic orbitals), and an active space of four electrons in four spatial orbitals [CAS(4,4)]. 
As a PQC ansatz for the active space state, we use the number-preserving fabric (NPF) ansatz introduced in \cite{anselmetti2021}, consisting of a fabric of spin-adapted orbital rotations and double excitations on sets of two spatial orbitals (four spin-orbitals). 
Four layers of this ansatz are enough to recover the exact CASCI ground state energy inside the active space of 4 orbitals, resulting in 20 PQC parameters. 
This is an overparameterization of the ground state, implying a global redundancy in the ansatz and resulting in a singular hessian at every point. 
The goal of this numerical demonstration is to show that Algorithm~\ref{alg:main} can still recover the Berry phase, in this case using regularization and backtracking.

\begin{figure}[t]
	\centering
    \includegraphics[width=0.48\textwidth]{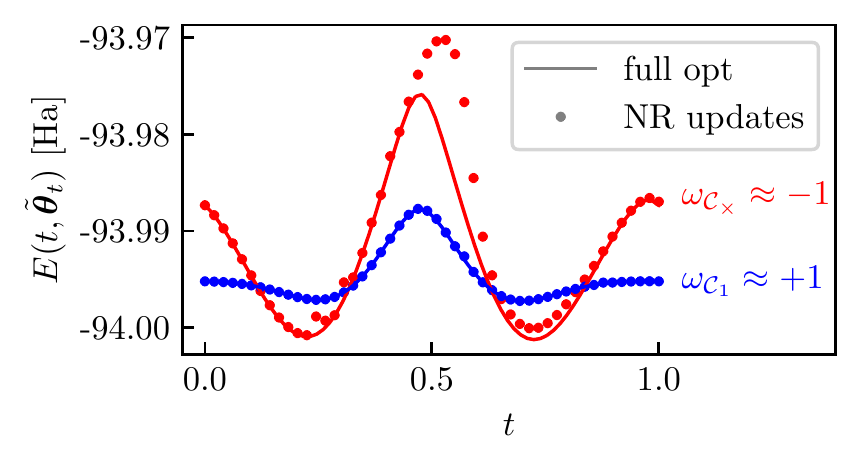}
    \caption{
        Energies throughout two loops, for a more challenging model of Formaldimine, described in the cc-pVDZ basis set and with a (4,4) active space.
        The (red) blue loop indicates a (non-)trivial Berry phase, centered around ($\alpha=\ang{113}$) $\alpha=\ang{145}$ and $\phi=\ang{90}$ (both), with a radius of $\ang{15}$, and discretized by $N=50$ points
        (note these loops are different from Fig.~\ref{fig:CI-resolution}).
        The basis set is too large to run a FCI calculation to locate the CI, and using another approximate method (e.g. state-average CASSCF) might bias the CI location.
        Instead, we manually choose larger loop geometries and use Algorithm~\ref{alg:main} to find the value of $\Pi_\C$.}
    \label{fig:energy_big_system}
\end{figure}

In Fig.~\ref{fig:energy_big_system} the energies throughout two loops are shown. 
The location of the conical intersection $(\alpha^\times, \phi^\times)$ in the larger basis set moves compared the case shown in Fig.~\ref{fig:CI-resolution} (this is to be expected, as the cc-pVDZ and STO-3G models are effectively different); the basis is now too large to attempt an FCI calculation that would resolve the gap exactly. 
One could instead resort to a State-Averaged CASSCF calculation to resolve the location of the CI, however, the state-average approach might bias the location of the CI. 
For this demonstration, we manually select two loops with a slightly larger radius of $\ang{15}$, centered around $\alpha = \ang{113}$ ($\C_\times$, red line) and around $\alpha = \ang{145}$ ($\C_1$, blue line). 
These values are chosen based on the location of the CI at the level of theory of large state-average CAS(14,14)SCF calculation, which returns $\alpha^\times \approx \ang{113}$.  
We choose $\phi = \ang{90}$, as the CI is forced to lie on the $\phi^\times = 90$ hyperplane due to the Cs reflection point-group symmetry. 
Due to the single-step Newton-Raphson optimization, we can observe a difference between the energies evaluated at each iteration point of our algorithm (circle markers) and the variational optima (solid lines).
The discrepancy is more pronounced after a large change in the state character, such as happens just after $t=0.5$ in the topologically-nontrivial loop $\C_\times$.
There, the variational parameters need to change more rapidly to track the state.
As long as the discretization step is small enough, the single Newton-Raphson steps ensure the variational state keeps tracking the local minimum.
Indeed, we can resolve the correct Berry phase with only $N=50$ discretiation points, recovering a final overlap [Eq.~\eqref{eq:final-overlap-with-G}] of $\omega_{\C_\times} = -0.9994$ (loop containing the CI), and $\omega_{\C_1} = 0.99998$ (trivial loop).

\section{Conclusion and outlook}
\label{sec:conclusion}

In this work, we introduced a hybrid algorithm to resolve conical intersections through the Berry phase they induce. 
This is achieved by tracking the ground state with a variational quantum ansatz, along a closed path $\C$ in nuclear configuration space. 
This algorithm only requires approximating the ground state (in contrast to e.g. the state-average VQE \cite{Yalouz2021}) for one nuclear geometry $\R$ at a time, reducing the expressivity requirements of the ansatz.
The key requirement of the algorithm is that the ansatz parameters are changed smoothly, tracking a local minimum of the cost function \eqref{eq:var_energy} and ensuring the $U(1)$-gauge (global phase) of the ansatz state remains well-defined.

As the output is quantized ($\Pi_\C \in \{0, \pi\}$), the result only needs to be estimated to a constant precision, and the algorithm is robust to some amount of error; we considered optimization error (the error on estimates of a variational parameter, due to the approximate minimizer) and sampling noise on the quantities measured on the quantum device.
We showed that one can update the variational parameters with a single newton step for each geometry in a discretization of the loop $\C$.
We proved analytically that the algorithm is granted to converge for a large enough number of discretization steps $N$, and small enough additive error, under the assumption that the cost function is strongly convex at its minimum. 
(We consider sampling noise explicitly, but the robustness results extends to any additive noise, including hardware noise.)
We argue this result practically extends to cases where the strong convexity assumption is not satisfied, as long as some regularization technique is employed.

This reasoning is corroborated by numerical demonstrations of CI resolution on a small example system -- the formaldimine molecule.
Using a minimal description of formaldimine [STO-3G basis, a CAS(2,2), UCCD ansatz], for which we have a strongly convex cost function, we show convergence of Algorithm~\ref{alg:main} without using regularization for a sufficiently large $N>11$, as we expect from our analytical results.
We also demonstrate the effect of sampling noise in this setting, showing that our algorithm is robust to a sizable amount of noise, achieving convergence for $N$ comparable to the noiseless case. 
Finally, we demonstrate the application of Algorithm~\ref{alg:main} with regularization on a more complicated and realistic model of formaldimine [cc-pVDZ basis, CAS(4,4), NPF ansatz]. 
This case shows that, even with a cost function that is never convex, we can employ regularization to resolve the Berry phase correctly.

\subsection{Paths towards improving convergence}
\label{sec:improvements}

The key step in our algorithm, where most of the cost in terms of quantum resources is concentrated, is the evaluation of the (NR) parameter update.
Ensuring the parameter estimates remain within the basin of convergence of the cost function is crucial, and it is the bottleneck in terms of the cost of our algorithm. 
As shown in Section~\ref{sec:error-bounding}, the size of the basin of convergence depends on the convexity of the cost function at the minimum \eqref{eq:strong-convexity-at-minimum}.
Overparametrizations (local or global) of the cost function are especially disrupting, as they produce singular Hessians ($m=0$) even at optimal points.
In this work, we proposed to use regularization by Hessian augmentation and back-tracking to solve this problem; this technique is practical and, as shown numerically in Section \ref{sec:larger-system}, it can produce convergent results for systems with overparametrized cost functions.
However, the success of these techniques may depend on the choice of their hyperparameters ($\alpha, \beta, \mu, \rho$ in Algorithm~\ref{alg:NR-regularized}), and their application makes the analytical study of the algorithm convergence harder.
In this section, we suggest alternative approaches to regularization and resource allocation for the algorithm, which would require further study.

\emph{Quantum natural gradients ---}
Quantum natural gradient (QNG) descent is a recently-proposed parameter update technique \cite{Stokes2020,Meyer2021}, which takes into account the geometry of the ground state manifold.
Natural gradient techniques, already long in use in classical machine learning \cite{amari1998}, are invariant with respect to reparametrizations of the cost function; more importantly for our case, they nullify the effect of overparametrizations \cite{liang2019}. 
The idea of QNG is to transform gradients with respect to the ansatz parameters into gradients with respect to Quantum Information Geometry.
This reparametrization is achieved through the Fubini-Study metric tensor $g(\thetaest_t)$ (to be evaluated at each update).
A gradient descent step would then become:
\begin{equation}
    \thetaest_{t+\Delta t} = \tilde{\bftheta}_{t} - \eta g^+(\thetaest_t) \boldsymbol{\mathcal{G}}(t+\Delta t, \thetaest_t),
\end{equation} 
where $\eta$ is the learning rate and $g^+(\thetaest_t)$ is the pseudo-inverse of the metric tensor. 
Here $\boldsymbol{\mathcal{G}}(t+\Delta t, \thetaest_t)$ is just the usual gradient of the energy as in Algorithm~\ref{alg:main}.
The resulting QNG step updates the ansatz by a fixed amount in the norm induced by the distance between quantum states, instead of a parameter-space norm, solving the issues connected to overparametrization.
Another option would be to use classical natural gradients (NG)~\cite{amari1998} defined on the 1- and 2-RDM manifold, which rely on the classical Fisher information matrix.

\emph{Adaptive step selection ---}
Choosing a sufficient number of steps $N$ to discretize $\C$ is key to the success of our algorithm, as proven in Section~\ref{sec:error-bounding} and shown in Fig.~\ref{fig:CI-resolution} (bottom right).
This is because the minimum $\thetaex_t$, along with its basin of convergence, changes between subsequent steps by an amount proportional to the step size $\Delta t = 1/N$.
In Algorithm~\ref{alg:main}, we propose to linearly discretize a given parametrization of $\C$ for the sake of simplicity.
An adaptive choice of $\Delta t$ could greatly reduce the cost of the algorithm, letting the steps be larger in the regions where the ground state changes the least, while concentrating more points in the regions where the ground state character sharply shifts.
An adaptive step selection technique that preserves the provable convergence could be easily implemented if the gradient and Hessian of the cost function are measured through the 1- and 2-electron RDM, as per Eq.~\eqref{eq:cost_rdm}.
The derivatives with respect to $\bftheta$ are then calculated by chain rule from the derivatives of the RDMs, at the current parameter value of $\bftheta=\thetaest_t$.
This allows to compute energy $E(t', \thetaest_t)$, gradient $\mathcal{G}(t', \thetaest_t)$ and Hessian $\mathcal{H}(t', \thetaest_t)$ for any value $t'$, without further evaluations of the PQC.
The step size can be then chosen as the maximum $\Delta t$ such that the Hessian convexity $m(t + \Delta t, \thetaest_t)$ remains above some positive threshold value.
Further research could quantify the improvement that adaptive step selection would bring to our algorithm, and identify a method to implement this for optimized energy derivatives sampling techniques, such as those using double factorization \cite{hohenstein2022}.

\subsection{Potential applications}

The algorithm we propose resolves the Berry phase along a given path $\C$; the description of the loop is an input of the algorithm.
This loop construction will depend on the details of the considered application, and might involve chemical intuition and the consideration symmetries of the molecule where present.
In realistic applications, we conceive our algorithm as a tool that can help to (1) certify CIs proposed by other methods, (2) determine whether a CI plays a role in a certain reaction, and/or (3) locate a point of the CI manifold in parameter space.
In either case, an initial proposal of a path $\C$ that might contain the CI is necessary.

The case 1 is the most direct application of our algorithm. 
The loop $\C$ is chosen to surround a {quasi-degeneracy} previously identified by an approximate classical method. 
The result of our algorithm could then confirm or disprove the presence of the CI.
In case 2, given a photochemical reaction whose geometry is approximately known, we can use our algorithm on a set of loops to understand wether a CI plays a role in the reaction.
These loops can be constructed by variations of the reaction path along perpendicular coordinates, focusing on the modes that influence the orbitals involved in the reaction, thus greatly reducing the search space for the CI.
Finally, to locate the CI (case 3) various search approaches can be considered.
For example, starting from a large loop $\C$ that is known to contain the CI, binary-search can be used to iteratively shrink the loop and locate the CI to the desired precision.
The considered loops could be defined on a plane, if there is heuristic information about the direction along which the potential energy surfaces split.
In alternative, a mesh of orthogonal loops in a subspace of the nuclear configuration space $\mathcal{R}$ can be tested.
This, in combination with the data about the ground state energy collected by running the optimization in our algorithm, could also be used to determine the approximate location of the minimal energy crossing point [i.e.~the point $\R_\text{MECP}$ on the CI manifold with minimum $E_0(\R_\text{MECP}) = E_1(\R_\text{MECP})$].
Further work is needed to explore these problems, develop procedures to solve them and define and test practical application cases in the three categories.

\subsection{Outlook}

The bounds presented in Section~\ref{sec:error-bounding} have been calculated to provide a guarantee of convergence for our method, which is an atypical feature for variational quantum algorithms.
These are not supposed to be tight bounds or resource estimates for a realistic application of our algorithm.
Further research is needed to define better bounds.
This, along with the choice of a specific method to extract the energy and its derivatives (e.g.~RDM sampling and parameter shift rule) could allow estimating the cost of a practical application of this algorithm.
Furthermore, a study of the errors due to the ansatz not perfectly reproducing the ground state, and those induced by circuit noise, could help to understand the practical limitations of the algorithm.

The computation of Berry phases is also central when characterizing topological phases of matter~\cite{asboth16}. 
For the specific case of non-interacting Hamiltonians with chiral or inversion symmetry, the winding number is analytically computed through the Zak phase~\cite{zak1989}, which is related to the Berry phase that is accumulated after a closed loop through the Brillouin Zone. 
For interacting systems, in which one cannot access momentum space, there is a mechanism in which one introduces an external periodic perturbation to the Hamiltonian~\cite{hatsugai2006, fukui2005}.
As long as the perturbation does not close the gap and respects the symmetries of the system, the Berry phase can be computed by considering a closed loop in parameter space, similar to what is proposed in this work. 
As an outlook, one could consider extending the VQE approach to detect topological phases of matter through the computation of the Zak phase. 
Furthermore, VQA approaches to other topological invariants, such as the Chern number \cite{chern1946} could be considered (possibly inspired by methods related to their experimental detection, such as Thouless pumping \cite{citro2023}).

Finally, classical algorithms to resolve conical intersections from Berry phases inspired by this approach could be designed.
If the approximate ground state is represented by a variational classical ansatz that fixes the $U(1)$ gauge, a simple extension of our method could be achievable. 
For example, this could be achieved with an extension of CASSCF (essentially a CASCI solver on top of an orbital optimization) that implements continuous local optimization of the SCF matrix, to allow enforcing the smoothness constrains that are crucial to keep the gauge of the state fixed.

\section*{Acknowledgements}
We thank Joonhoo Lee, Bill Huggins, Nick Rubin, Ryan Babbush, Franco Buda and Maciej Lewenstein for stimulating discussions.
We thank Dyon van Vreumingen and Patrick Emonts for their useful feedback.

EK and SP acknowledge support from Shell Global Solutions BV. 
AD acknowledges funding from the European Union under Grant Agreement 101080142 and the project EQUALITY.
ICFO group acknowledges support from: ERC AdG NOQIA; Ministerio de Ciencia y Innovation Agencia Estatal de Investigaciones (PGC2018-097027-B-100/10.13039/501100011033, CEX2019-000910-S/10.13039/501100011033, Plan National FIDEUA PID2019-106901GB-I00, FPI, QUANTERA MAQS PCI2019-111828-2, QUANTERA DYNAMITE PCI2022-132919,  Proyectos de I+D+I “Retos Colaboraci\'on” QUSPIN RTC2019-007196-7); MICIIN with funding from European Union NextGenerationEU(PRTR-C17.I1) and by Generalitat de Catalunya;  Fundació Cellex; Fundació Mir-Puig; Generalitat de Catalunya (European social fund FEDER and CERCA program, AGAUR Grant No. 2021 SGR 01452, QuantumCAT / U16-011424, co-funded by ERDF Operational Program of Catalonia 2014-2020); Barcelona Supercomputing Center MareNostrum (FI-2022-1-0042); EU Horizon 2020 FET-OPEN OPTOlogic (Grant No 899794); EU Horizon Europe Program (Grant Agreement 101080086 — NeQST), National Science Centre, Poland (Symfonia Grant No. 2016/20/W/ST4/00314); ICFO Internal “QuantumGaudi” project; European Union’s Horizon 2020 research and innovation program under the Marie-Skłodowska-Curie grant agreement No 101029393 (STREDCH) and No 847648  (“La Caixa” Junior Leaders fellowships ID100010434: LCF/BQ/PI19/11690013, LCF/BQ/PI20/11760031,  LCF/BQ/PR20/11770012, LCF/BQ/PR21/11840013). Views and opinions expressed in this work are, however, those of the author(s) only and do not necessarily reflect those of the European Union, European Climate, Infrastructure and Environment Executive Agency (CINEA), nor any other granting authority.

\bibliographystyle{unsrtnat}
\bibliography{biblio}

\providecommand{\noopsort}[1]{}
\begin{thebibliography}{78}
\providecommand{\natexlab}[1]{#1}
\providecommand{\url}[1]{\texttt{#1}}
\expandafter\ifx\csname urlstyle\endcsname\relax
  \providecommand{\doi}[1]{doi: #1}\else
  \providecommand{\doi}{doi: \begingroup \urlstyle{rm}\Url}\fi

\bibitem[Geim and Novoselov(2007)]{geim2007}
A.~K. Geim and K.~S. Novoselov.
\newblock The rise of graphene.
\newblock \emph{Nature Materials}, 6\penalty0 (3):\penalty0 183--191, March
  2007.
\newblock ISSN 1476-4660.
\newblock \doi{10.1038/nmat1849}.

\bibitem[Berry(1984)]{berry1984}
Michael~Victor Berry.
\newblock Quantal phase factors accompanying adiabatic changes.
\newblock \emph{Proceedings of the Royal Society of London. A. Mathematical and
  Physical Sciences}, 392\penalty0 (1802):\penalty0 45--57, March 1984.
\newblock \doi{10.1098/rspa.1984.0023}.

\bibitem[Domcke et~al.(2011)Domcke, Yarkony, and K{\"o}ppel]{domcke2011}
Wolfgang Domcke, David Yarkony, and Horst K{\"o}ppel, editors.
\newblock \emph{Conical Intersections: Theory, Computation and Experiment}.
\newblock Number v. 17 in Advanced Series in Physical Chemistry. {World
  Scientific}, {Singapore ; Hackensack, NJ}, 2011.
\newblock ISBN 978-981-4313-44-5.

\bibitem[Yarkony(2012)]{yarkony2012}
David~R. Yarkony.
\newblock Nonadiabatic {{Quantum Chemistry}}\textemdash{{Past}}, {{Present}},
  and {{Future}}.
\newblock \emph{Chemical Reviews}, 112\penalty0 (1):\penalty0 481--498, January
  2012.
\newblock ISSN 0009-2665.
\newblock \doi{10.1021/cr2001299}.

\bibitem[Polli et~al.(2010)Polli, Alto{\`e}, Weingart, Spillane, Manzoni,
  Brida, Tomasello, Orlandi, Kukura, Mathies, Garavelli, and
  Cerullo]{polli2010}
Dario Polli, Piero Alto{\`e}, Oliver Weingart, Katelyn~M. Spillane, Cristian
  Manzoni, Daniele Brida, Gaia Tomasello, Giorgio Orlandi, Philipp Kukura,
  Richard~A. Mathies, Marco Garavelli, and Giulio Cerullo.
\newblock Conical intersection dynamics of the primary photoisomerization event
  in vision.
\newblock \emph{Nature}, 467\penalty0 (7314):\penalty0 440--443, September
  2010.
\newblock ISSN 1476-4687.
\newblock \doi{10.1038/nature09346}.

\bibitem[{Olaso-Gonz{\'a}lez} et~al.(2006){Olaso-Gonz{\'a}lez}, Merch{\'a}n,
  and {Serrano-Andr{\'e}s}]{olaso-gonzalez2006}
Gloria {Olaso-Gonz{\'a}lez}, Manuela Merch{\'a}n, and Luis
  {Serrano-Andr{\'e}s}.
\newblock Ultrafast {{Electron Transfer}} in {{Photosynthesis}}: {{Reduced
  Pheophytin}} and {{Quinone Interaction Mediated}} by {{Conical
  Intersections}}.
\newblock \emph{The Journal of Physical Chemistry B}, 110\penalty0
  (48):\penalty0 24734--24739, December 2006.
\newblock ISSN 1520-6106, 1520-5207.
\newblock \doi{10.1021/jp063915u}.

\bibitem[Zimmerman(1966)]{Zimmerman1966}
Howard~E Zimmerman.
\newblock {Molecular Orbital Correlation Diagrams, Mobius Systems, and Factors
  Controlling Ground- and Excited-State Reactions. II}.
\newblock \emph{Journal of the American Chemical Society}, 88\penalty0
  (7):\penalty0 1566--1567, 1966.
\newblock ISSN 0002-7863.
\newblock \doi{10.1021/ja00959a053}.

\bibitem[Bernardi et~al.(1996)Bernardi, Olivucci, and Robb]{Bernardi1996}
Fernando Bernardi, Massimo Olivucci, and Michael~A. Robb.
\newblock {Potential energy surface crossings in organic photochemistry}.
\newblock \emph{Chemical Society Reviews}, 25\penalty0 (5):\penalty0 321--328,
  1996.
\newblock ISSN 0306-0012.
\newblock \doi{10.1039/cs9962500321}.

\bibitem[González et~al.(2012)González, Escudero, and
  Serrano‐Andrés]{Gonzalez2012}
Leticia González, Daniel Escudero, and Luis Serrano‐Andrés.
\newblock {Progress and Challenges in the Calculation of Electronic Excited
  States}.
\newblock \emph{ChemPhysChem}, 13\penalty0 (1):\penalty0 28--51, 2012.
\newblock ISSN 1439-4235.
\newblock \doi{10.1002/cphc.201100200}.

\bibitem[Feynman(1982)]{feynman1982}
Richard~P. Feynman.
\newblock Simulating physics with computers.
\newblock \emph{International Journal of Theoretical Physics}, 21\penalty0
  (6-7):\penalty0 467--488, June 1982.
\newblock ISSN 0020-7748, 1572-9575.
\newblock \doi{10.1007/BF02650179}.

\bibitem[{Aspuru-Guzik} et~al.(2005){Aspuru-Guzik}, Dutoi, Love, and
  {Head-Gordon}]{aspuru-guzik2005}
Al{\'a}n {Aspuru-Guzik}, Anthony~D. Dutoi, Peter~J. Love, and Martin
  {Head-Gordon}.
\newblock Simulated {{Quantum Computation}} of {{Molecular Energies}}.
\newblock \emph{Science}, 309\penalty0 (5741):\penalty0 1704--1707, September
  2005.
\newblock \doi{10.1126/science.1113479}.

\bibitem[Preskill(2018)]{preskill2018}
John Preskill.
\newblock Quantum {{Computing}} in the {{NISQ}} era and beyond.
\newblock \emph{Quantum}, 2:\penalty0 79, August 2018.
\newblock ISSN 2521-327X.
\newblock \doi{10.22331/q-2018-08-06-79}.

\bibitem[Peruzzo et~al.(2014)Peruzzo, McClean, Shadbolt, Yung, Zhou, Love,
  {Aspuru-Guzik}, and O'Brien]{peruzzo2014}
Alberto Peruzzo, Jarrod~R. McClean, Peter Shadbolt, Man-Hong Yung, Xiao-Qi
  Zhou, Peter~J. Love, Al{\'a}n {Aspuru-Guzik}, and Jeremy~L. O'Brien.
\newblock A variational eigenvalue solver on a photonic quantum processor.
\newblock \emph{Nature Communications}, 5\penalty0 (1):\penalty0 4213,
  September 2014.
\newblock ISSN 2041-1723.
\newblock \doi{10.1038/ncomms5213}.

\bibitem[McClean et~al.(2016)McClean, Romero, Babbush, and
  {Aspuru-Guzik}]{mcclean2016}
Jarrod~R. McClean, Jonathan Romero, Ryan Babbush, and Al{\'a}n {Aspuru-Guzik}.
\newblock The theory of variational hybrid quantum-classical algorithms.
\newblock \emph{New Journal of Physics}, 18\penalty0 (2):\penalty0 023023,
  February 2016.
\newblock ISSN 1367-2630.
\newblock \doi{10.1088/1367-2630/18/2/023023}.

\bibitem[Wecker et~al.(2015)Wecker, Hastings, and Troyer]{wecker2015}
Dave Wecker, Matthew~B Hastings, and Matthias Troyer.
\newblock Progress towards practical quantum variational algorithms.
\newblock \emph{Physical Review A}, 92\penalty0 (4):\penalty0 042303, October
  2015.
\newblock ISSN 1050-2947.
\newblock \doi{10.1103/PhysRevA.92.042303}.

\bibitem[McClean et~al.(2018)McClean, Boixo, Smelyanskiy, Babbush, and
  Neven]{mcclean2018}
Jarrod~R. McClean, Sergio Boixo, Vadim~N. Smelyanskiy, Ryan Babbush, and
  Hartmut Neven.
\newblock Barren plateaus in quantum neural network training landscapes.
\newblock \emph{Nature Communications}, 9\penalty0 (1):\penalty0 4812, November
  2018.
\newblock ISSN 2041-1723.
\newblock \doi{10.1038/s41467-018-07090-4}.

\bibitem[Tamiya et~al.(2021)Tamiya, Koh, and Nakagawa]{tamiya2020}
Shiro Tamiya, Sho Koh, and Yuya~O. Nakagawa.
\newblock Calculating nonadiabatic couplings and berry's phase by variational
  quantum eigensolvers.
\newblock \emph{Phys. Rev. Research}, 3:\penalty0 023244, Jun 2021.
\newblock \doi{10.1103/PhysRevResearch.3.023244}.

\bibitem[Xiao et~al.(2022)Xiao, Freericks, and Kemper]{xiao2022}
Xiao Xiao, J.~K. Freericks, and A.~F. Kemper.
\newblock Robust measurement of wave function topology on {{NISQ}} quantum
  computers, October 2022.
\newblock URL \url{https://doi.10.48550/arXiv.2101.07283}.

\bibitem[Murta et~al.(2020)Murta, Catarina, and Fern\'andez-Rossier]{murta2020}
Bruno Murta, G.~Catarina, and J.~Fern\'andez-Rossier.
\newblock Berry phase estimation in gate-based adiabatic quantum simulation.
\newblock \emph{Phys. Rev. A}, 101:\penalty0 020302, Feb 2020.
\newblock \doi{10.1103/PhysRevA.101.020302}.
\newblock URL \url{https://link.aps.org/doi/10.1103/PhysRevA.101.020302}.

\bibitem[{Longuet-Higgins} et~al.(1958){Longuet-Higgins}, {\"O}pik, Pryce, and
  Sack]{longuet-higgins1958}
Hugh~Christopher {Longuet-Higgins}, U.~{\"O}pik, Maurice Henry~Lecorney Pryce,
  and R.~A. Sack.
\newblock Studies of the {{Jahn-Teller}} effect .{{II}}. {{The}} dynamical
  problem.
\newblock \emph{Proceedings of the Royal Society of London. Series A.
  Mathematical and Physical Sciences}, 244\penalty0 (1236):\penalty0 1--16,
  February 1958.
\newblock \doi{10.1098/rspa.1958.0022}.

\bibitem[Mead and Truhlar(1979)]{mead1979a}
C.~Alden Mead and Donald~G. Truhlar.
\newblock On the determination of {{Born}}\textendash{{Oppenheimer}} nuclear
  motion wave functions including complications due to conical intersections
  and identical nuclei.
\newblock \emph{The Journal of Chemical Physics}, 70\penalty0 (5):\penalty0
  2284--2296, March 1979.
\newblock ISSN 0021-9606.
\newblock \doi{10.1063/1.437734}.

\bibitem[Ryabinkin et~al.(2017)Ryabinkin, {Joubert-Doriol}, and
  Izmaylov]{ryabinkin2017}
Ilya~G. Ryabinkin, Lo{\"i}c {Joubert-Doriol}, and Artur~F. Izmaylov.
\newblock Geometric {{Phase Effects}} in {{Nonadiabatic Dynamics}} near
  {{Conical Intersections}}.
\newblock \emph{Accounts of Chemical Research}, 50\penalty0 (7):\penalty0
  1785--1793, July 2017.
\newblock ISSN 0001-4842.
\newblock \doi{10.1021/acs.accounts.7b00220}.

\bibitem[Whitlow et~al.(2023)Whitlow, Jia, Wang, Fang, Kim, and
  Brown]{whitlow2023}
Jacob Whitlow, Zhubing Jia, Ye~Wang, Chao Fang, Jungsang Kim, and Kenneth~R.
  Brown.
\newblock Simulating conical intersections with trapped ions, February 2023.
\newblock URL \url{https://doi.org/10.48550/arXiv.2211.07319}.

\bibitem[Valahu et~al.(2023)Valahu, {Olaya-Agudelo}, MacDonell, Navickas, Rao,
  Millican, {P{\'e}rez-S{\'a}nchez}, {Yuen-Zhou}, Biercuk, Hempel, Tan, and
  Kassal]{valahu2023}
Christophe~H. Valahu, Vanessa~C. {Olaya-Agudelo}, Ryan~J. MacDonell, Tomas
  Navickas, Arjun~D. Rao, Maverick~J. Millican, Juan~B.
  {P{\'e}rez-S{\'a}nchez}, Joel {Yuen-Zhou}, Michael~J. Biercuk, Cornelius
  Hempel, Ting~Rei Tan, and Ivan Kassal.
\newblock Direct observation of geometric phase in dynamics around a conical
  intersection.
\newblock \emph{Nature Chemistry}, 15\penalty0 (11):\penalty0 1503--1508,
  November 2023.
\newblock ISSN 1755-4330, 1755-4349.
\newblock \doi{10.1038/s41557-023-01300-3}.

\bibitem[Wang et~al.(2023)Wang, Frattini, Chapman, Puri, Girvin, Devoret, and
  Schoelkopf]{wang2023}
Christopher~S. Wang, Nicholas~E. Frattini, Benjamin~J. Chapman, Shruti Puri,
  Steven~M. Girvin, Michel~H. Devoret, and Robert~J. Schoelkopf.
\newblock Observation of wave-packet branching through an engineered conical
  intersection.
\newblock \emph{Physical Review X}, 13\penalty0 (1):\penalty0 011008, January
  2023.
\newblock ISSN 2160-3308.
\newblock \doi{10.1103/PhysRevX.13.011008}.

\bibitem[Koridon and Polla(2024)]{auto_oo}
Emiel Koridon and Stefano Polla.
\newblock auto\_oo: an autodifferentiable framework for molecular
  orbital-optimized variational quantum algorithms.
\newblock Zenodo, February 2024.
\newblock URL \url{https://doi.org/10.5281/zenodo.10639817}.

\bibitem[Teller(1937)]{teller1937}
E.~Teller.
\newblock The {{Crossing}} of {{Potential Surfaces}}.
\newblock \emph{The Journal of Physical Chemistry}, 41\penalty0 (1):\penalty0
  109--116, January 1937.
\newblock ISSN 0092-7325.
\newblock \doi{10.1021/j150379a010}.

\bibitem[Herzberg and {Longuet-Higgins}(1963)]{herzberg1963}
G.~Herzberg and H.~C. {Longuet-Higgins}.
\newblock Intersection of potential energy surfaces in polyatomic molecules.
\newblock \emph{Discussions of the Faraday Society}, 35\penalty0 (0):\penalty0
  77--82, January 1963.
\newblock ISSN 0366-9033.
\newblock \doi{10.1039/DF9633500077}.

\bibitem[Helgaker et~al.(2000)Helgaker, J{\o}rgensen, and Olsen]{helgaker2000}
Trygve Helgaker, Poul J{\o}rgensen, and Jeppe Olsen.
\newblock \emph{Molecular {{Electronic}}-{{Structure Theory}}}.
\newblock {Wiley}, first edition, August 2000.
\newblock ISBN 978-0-471-96755-2 978-1-119-01957-2.
\newblock \doi{10.1002/9781119019572}.

\bibitem[Broer et~al.(2003)Broer, Hozoi, and Nieuwpoort]{broer2003}
R.~Broer, L.~Hozoi, and W.~C. Nieuwpoort.
\newblock Non-orthogonal approaches to the study of magnetic interactions.
\newblock \emph{Molecular Physics}, 101\penalty0 (1-2):\penalty0 233--240,
  January 2003.
\newblock ISSN 0026-8976.
\newblock \doi{10.1080/0026897021000035205}.

\bibitem[Veryazov et~al.(2011)Veryazov, Malmqvist, and Roos]{veryazov2011}
Valera Veryazov, Per~{\AA}ke Malmqvist, and Bj{\"o}rn~O. Roos.
\newblock How to select active space for multiconfigurational quantum
  chemistry?
\newblock \emph{International Journal of Quantum Chemistry}, 111\penalty0
  (13):\penalty0 3329--3338, 2011.
\newblock ISSN 1097-461X.
\newblock \doi{10.1002/qua.23068}.

\bibitem[Yarkony(1996)]{Yarkony1996}
David~R. Yarkony.
\newblock Diabolical conical intersections.
\newblock \emph{Reviews of Modern Physics}, 68\penalty0 (4):\penalty0
  985--1013, October 1996.
\newblock \doi{10.1103/RevModPhys.68.985}.

\bibitem[Alden~Mead(1980)]{aldenmead1980}
C.~Alden~Mead.
\newblock The molecular {{Aharonov}}\textemdash{{Bohm}} effect in bound states.
\newblock \emph{Chemical Physics}, 49\penalty0 (1):\penalty0 23--32, June 1980.
\newblock ISSN 0301-0104.
\newblock \doi{10.1016/0301-0104(80)85035-X}.

\bibitem[Harwood et~al.(2022)Harwood, Trenev, Stober, Barkoutsos, Gujarati,
  Mostame, and Greenberg]{harwood2022}
Stuart~M. Harwood, Dimitar Trenev, Spencer~T. Stober, Panagiotis Barkoutsos,
  Tanvi~P. Gujarati, Sarah Mostame, and Donny Greenberg.
\newblock Improving the {{Variational Quantum Eigensolver Using Variational
  Adiabatic Quantum Computing}}.
\newblock \emph{ACM Transactions on Quantum Computing}, 3\penalty0
  (1):\penalty0 1:1--1:20, January 2022.
\newblock ISSN 2643-6809.
\newblock \doi{10.1145/3479197}.

\bibitem[Mead(1979)]{mead1979}
C.~Alden Mead.
\newblock The ''noncrossing'' rule for electronic potential energy surfaces:
  {{The}} role of time-reversal invariance.
\newblock \emph{The Journal of Chemical Physics}, 70\penalty0 (5):\penalty0
  2276--2283, March 1979.
\newblock ISSN 0021-9606.
\newblock \doi{10.1063/1.437733}.

\bibitem[Bartlett et~al.(1989)Bartlett, Kucharski, and Noga]{bartlett1989}
Rodney~J. Bartlett, Stanislaw~A. Kucharski, and Jozef Noga.
\newblock Alternative coupled-cluster ans\"atze {{II}}. {{The}} unitary
  coupled-cluster method.
\newblock \emph{Chemical Physics Letters}, 155\penalty0 (1):\penalty0 133--140,
  February 1989.
\newblock ISSN 0009-2614.
\newblock \doi{10.1016/S0009-2614(89)87372-5}.

\bibitem[Romero et~al.(2018)Romero, Babbush, McClean, Hempel, Love, and
  {Aspuru-Guzik}]{romero2018}
Jonathan Romero, Ryan Babbush, Jarrod~R. McClean, Cornelius Hempel, Peter~J.
  Love, and Al{\'a}n {Aspuru-Guzik}.
\newblock Strategies for quantum computing molecular energies using the unitary
  coupled cluster ansatz.
\newblock \emph{Quantum Science and Technology}, 4\penalty0 (1):\penalty0
  014008, October 2018.
\newblock ISSN 2058-9565.
\newblock \doi{10.1088/2058-9565/aad3e4}.

\bibitem[Anselmetti et~al.(2021)Anselmetti, Wierichs, Gogolin, and
  Parrish]{anselmetti2021}
Gian-Luca~R. Anselmetti, David Wierichs, Christian Gogolin, and Robert~M.
  Parrish.
\newblock Local, expressive, quantum-number-preserving vqe ansatze for
  fermionic systems.
\newblock \emph{New Journal of Physics}, 23, 4 2021.
\newblock \doi{10.1088/1367-2630/ac2cb3}.

\bibitem[Schuld et~al.(2019)Schuld, Bergholm, Gogolin, Izaac, and
  Killoran]{schuld2019}
Maria Schuld, Ville Bergholm, Christian Gogolin, Josh Izaac, and Nathan
  Killoran.
\newblock Evaluating analytic gradients on quantum hardware.
\newblock \emph{Physical Review A}, 99\penalty0 (3):\penalty0 032331, March
  2019.
\newblock ISSN 2469-9926, 2469-9934.
\newblock \doi{10.1103/PhysRevA.99.032331}.

\bibitem[Jensen and Jorgensen(1984)]{jensen1984}
Hans Jorgen~Aa. Jensen and Poul Jorgensen.
\newblock A direct approach to second-order {{MCSCF}} calculations using a norm
  extended optimization scheme.
\newblock \emph{The Journal of Chemical Physics}, 80\penalty0 (3):\penalty0
  1204--1214, February 1984.
\newblock ISSN 0021-9606.
\newblock \doi{10.1063/1.446797}.

\bibitem[{Helmich-Paris}(2021)]{helmich-paris2021}
Benjamin {Helmich-Paris}.
\newblock A trust-region augmented {{Hessian}} implementation for restricted
  and unrestricted {{Hartree}}\textendash{{Fock}} and
  {{Kohn}}\textendash{{Sham}} methods.
\newblock \emph{The Journal of Chemical Physics}, 154\penalty0 (16):\penalty0
  164104, April 2021.
\newblock ISSN 0021-9606.
\newblock \doi{10.1063/5.0040798}.

\bibitem[O'Brien et~al.(2021)O'Brien, Polla, Rubin, Huggins, McArdle, Boixo,
  McClean, and Babbush]{OBrien2021}
Thomas~E. O'Brien, Stefano Polla, Nicholas~C. Rubin, William~J. Huggins, Sam
  McArdle, Sergio Boixo, Jarrod~R. McClean, and Ryan Babbush.
\newblock {Error Mitigation via Verified Phase Estimation}.
\newblock \emph{PRX Quantum}, 2\penalty0 (2), oct 2021.
\newblock \doi{10.1103/prxquantum.2.020317}.

\bibitem[Polla et~al.(2023)Polla, Anselmetti, and O'Brien]{Polla2023}
Stefano Polla, Gian-Luca~R. Anselmetti, and Thomas~E. O'Brien.
\newblock Optimizing the information extracted by a single qubit measurement.
\newblock \emph{Physical Review A}, 108\penalty0 (1):\penalty0 012403, July
  2023.
\newblock \doi{10.1103/PhysRevA.108.012403}.

\bibitem[Nocedal and Wright(2006)]{nocedal2006}
Jorge Nocedal and Stephen~J. Wright.
\newblock \emph{Numerical Optimization}.
\newblock Springer Series in Operations Research. {Springer}, {New York}, 2nd
  ed edition, 2006.
\newblock ISBN 978-0-387-30303-1.

\bibitem[Wigner(1955)]{wigner1955}
Eugene~P. Wigner.
\newblock Characteristic {{Vectors}} of {{Bordered Matrices With Infinite
  Dimensions}}.
\newblock \emph{Annals of Mathematics}, 62\penalty0 (3):\penalty0 548--564,
  1955.
\newblock ISSN 0003-486X.
\newblock \doi{10.2307/1970079}.

\bibitem[Yalouz et~al.(2021)Yalouz, Senjean, G{\"{u}}nther, Buda, O'Brien, and
  Visscher]{Yalouz2021}
Saad Yalouz, Bruno Senjean, Jakob G{\"{u}}nther, Francesco Buda, Thomas~E
  O'Brien, and Lucas Visscher.
\newblock {A state-averaged orbital-optimized hybrid quantum–classical
  algorithm for a democratic description of ground and excited states}.
\newblock \emph{Quantum Science and Technology}, 6\penalty0 (2):\penalty0
  024004, jan 2021.
\newblock ISSN 2058-9565.
\newblock \doi{10.1088/2058-9565/abd334}.

\bibitem[Yalouz et~al.(2022)Yalouz, Koridon, Senjean, Lasorne, Buda, and
  Visscher]{yalouz2022analytical}
Saad Yalouz, Emiel Koridon, Bruno Senjean, Benjamin Lasorne, Francesco Buda,
  and Lucas Visscher.
\newblock Analytical nonadiabatic couplings and gradients within the
  state-averaged orbital-optimized variational quantum eigensolver.
\newblock \emph{Journal of Chemical Theory and Computation}, 18\penalty0
  (2):\penalty0 776--794, 2022.
\newblock \doi{10.1021/acs.jctc.1c00995}.
\newblock PMID: 35029988.

\bibitem[Löwdin(1950)]{lowdin1950non}
Per‐Olov Löwdin.
\newblock On the non‐orthogonality problem connected with the use of atomic
  wave functions in the theory of molecules and crystals.
\newblock \emph{The Journal of Chemical Physics}, 18\penalty0 (3):\penalty0
  365--375, 1950.
\newblock \doi{10.1063/1.1747632}.

\bibitem[{Bonet-Monroig} et~al.(2020){Bonet-Monroig}, Babbush, and
  O'Brien]{bonet-monroig2020}
Xavier {Bonet-Monroig}, Ryan Babbush, and Thomas~E. O'Brien.
\newblock Nearly {{Optimal Measurement Scheduling}} for {{Partial Tomography}}
  of {{Quantum States}}.
\newblock \emph{Physical Review X}, 10\penalty0 (3):\penalty0 031064, September
  2020.
\newblock \doi{10.1103/PhysRevX.10.031064}.

\bibitem[{\noopsort{burg}}{von Burg} et~al.(2021){\noopsort{burg}}{von Burg},
  Low, H{\"a}ner, Steiger, Reiher, Roetteler, and Troyer]{vonburg2021}
Vera {\noopsort{burg}}{von Burg}, Guang~Hao Low, Thomas H{\"a}ner, Damian~S.
  Steiger, Markus Reiher, Martin Roetteler, and Matthias Troyer.
\newblock Quantum computing enhanced computational catalysis.
\newblock \emph{Physical Review Research}, 3\penalty0 (3):\penalty0 033055,
  July 2021.
\newblock ISSN 2643-1564.
\newblock \doi{10.1103/PhysRevResearch.3.033055}.

\bibitem[Cohn et~al.(2021)Cohn, Motta, and Parrish]{cohn2021}
Jeffrey Cohn, Mario Motta, and Robert~M. Parrish.
\newblock Quantum {{Filter Diagonalization}} with {{Compressed
  Double-Factorized Hamiltonians}}.
\newblock \emph{PRX Quantum}, 2\penalty0 (4):\penalty0 040352, December 2021.
\newblock \doi{10.1103/PRXQuantum.2.040352}.

\bibitem[Arute et~al.(2020)Arute, Arya, Babbush, Bacon, Bardin, Barends, Boixo,
  Broughton, Buckley, Buell, Burkett, Bushnell, Chen, Chen, Chiaro, Collins,
  Courtney, Demura, Dunsworth, Farhi, Fowler, Foxen, Gidney, Giustina, Graff,
  Habegger, Harrigan, Ho, Hong, Huang, Huggins, Ioffe, Isakov, Jeffrey, Jiang,
  Jones, Kafri, Kechedzhi, Kelly, Kim, Klimov, Korotkov, Kostritsa, Landhuis,
  Laptev, Lindmark, Lucero, Martin, Martinis, McClean, McEwen, Megrant, Mi,
  Mohseni, Mruczkiewicz, Mutus, Naaman, Neeley, Neill, Neven, Niu, O'Brien,
  Ostby, Petukhov, Putterman, Quintana, Roushan, Rubin, Sank, Satzinger,
  Smelyanskiy, Strain, Sung, Szalay, Takeshita, Vainsencher, White, Wiebe, Yao,
  Yeh, and Zalcman]{arute2020}
Frank Arute, Kunal Arya, Ryan Babbush, Dave Bacon, Joseph~C. Bardin, Rami
  Barends, Sergio Boixo, Michael Broughton, Bob~B. Buckley, David~A. Buell,
  Brian Burkett, Nicholas Bushnell, Yu~Chen, Zijun Chen, Benjamin Chiaro,
  Roberto Collins, William Courtney, Sean Demura, Andrew Dunsworth, Edward
  Farhi, Austin Fowler, Brooks Foxen, Craig Gidney, Marissa Giustina, Rob
  Graff, Steve Habegger, Matthew~P. Harrigan, Alan Ho, Sabrina Hong, Trent
  Huang, William~J Huggins, Lev Ioffe, Sergei~V. Isakov, Evan Jeffrey, Zhang
  Jiang, Cody Jones, Dvir Kafri, Kostyantyn Kechedzhi, Julian Kelly, Seon Kim,
  Paul~V. Klimov, Alexander Korotkov, Fedor Kostritsa, David Landhuis, Pavel
  Laptev, Mike Lindmark, Erik Lucero, Orion Martin, John~M. Martinis, Jarrod~R.
  McClean, Matt McEwen, Anthony Megrant, Xiao Mi, Masoud Mohseni, Wojciech
  Mruczkiewicz, Josh Mutus, Ofer Naaman, Matthew Neeley, Charles Neill, Hartmut
  Neven, Murphy~Yuezhen Niu, Thomas~E. O'Brien, Eric Ostby, Andre Petukhov,
  Harald Putterman, Chris Quintana, Pedram Roushan, Nicholas~C. Rubin, Daniel
  Sank, Kevin~J. Satzinger, Vadim Smelyanskiy, Doug Strain, Kevin~J. Sung,
  Marco Szalay, Tyler~Y. Takeshita, Amit Vainsencher, Theodore White, Nathan
  Wiebe, Z.~Jamie Yao, Ping Yeh, and Adam Zalcman.
\newblock Hartree-{{Fock}} on a superconducting qubit quantum computer.
\newblock \emph{Science}, 369\penalty0 (6507):\penalty0 1084--1089, August
  2020.
\newblock ISSN 0036-8075.
\newblock \doi{10.1126/science.abb9811}.

\bibitem[Huembeli and Dauphin(2021)]{huembeli2021}
Patrick Huembeli and Alexandre Dauphin.
\newblock Characterizing the loss landscape of variational quantum circuits.
\newblock \emph{Quantum Science and Technology}, 6\penalty0 (2):\penalty0
  025011, February 2021.
\newblock ISSN 2058-9565.
\newblock \doi{10.1088/2058-9565/abdbc9}.

\bibitem[Hirai(2022)]{hirai2022}
Hirotoshi Hirai.
\newblock Excited-state molecular dynamics simulation based on variational
  quantum algorithms, November 2022.
\newblock URL \url{https://doi.org/10.48550/arXiv.2211.02302}.

\bibitem[{Bona{\v c}i{\'c}-Kouteck{\'y}} and Michl(1985)]{bonacic-koutecky1985}
Vlasta {Bona{\v c}i{\'c}-Kouteck{\'y}} and Josef Michl.
\newblock Photochemicalsyn-anti isomerization of a {{Schiff}} base: {{A}}
  two-dimensional description of a conical intersection in formaldimine.
\newblock \emph{Theoretica chimica acta}, 68\penalty0 (1):\penalty0 45--55,
  July 1985.
\newblock ISSN 1432-2234.
\newblock \doi{10.1007/BF00698750}.

\bibitem[Birge(1990)]{birge1990}
Robert~R. Birge.
\newblock Nature of the primary photochemical events in rhodopsin and
  bacteriorhodopsin.
\newblock \emph{Biochimica et Biophysica Acta (BBA) - Bioenergetics},
  1016\penalty0 (3):\penalty0 293--327, April 1990.
\newblock ISSN 0005-2728.
\newblock \doi{10.1016/0005-2728(90)90163-X}.

\bibitem[Chahre(1985)]{chahre1985}
M~Chahre.
\newblock Trigger and {{Amplification Mechanisms}} in {{Visual
  Phototransduction}}.
\newblock \emph{Annual Review of Biophysics and Biophysical Chemistry},
  14\penalty0 (1):\penalty0 331--360, 1985.
\newblock \doi{10.1146/annurev.bb.14.060185.001555}.

\bibitem[Bergholm et~al.(2022)Bergholm, Izaac, Schuld, Gogolin, Ahmed, Ajith,
  Alam, {Alonso-Linaje}, AkashNarayanan, Asadi, Arrazola, Azad, Banning, Blank,
  Bromley, Cordier, Ceroni, Delgado, Di~Matteo, Dusko, Garg, Guala, Hayes,
  Hill, Ijaz, Isacsson, Ittah, Jahangiri, Jain, Jiang, Khandelwal, Kottmann,
  Lang, Lee, Loke, Lowe, McKiernan, Meyer, {Monta{\~n}ez-Barrera}, Moyard, Niu,
  O'Riordan, Oud, Panigrahi, Park, Polatajko, Quesada, Roberts, S{\'a}, Schoch,
  Shi, Shu, Sim, Singh, Strandberg, Soni, Sz{\'a}va, Thabet,
  {Vargas-Hern{\'a}ndez}, Vincent, Vitucci, Weber, Wierichs, Wiersema,
  Willmann, Wong, Zhang, and Killoran]{bergholm2022}
Ville Bergholm, Josh Izaac, Maria Schuld, Christian Gogolin, Shahnawaz Ahmed,
  Vishnu Ajith, M.~Sohaib Alam, Guillermo {Alonso-Linaje}, B.~AkashNarayanan,
  Ali Asadi, Juan~Miguel Arrazola, Utkarsh Azad, Sam Banning, Carsten Blank,
  Thomas~R. Bromley, Benjamin~A. Cordier, Jack Ceroni, Alain Delgado, Olivia
  Di~Matteo, Amintor Dusko, Tanya Garg, Diego Guala, Anthony Hayes, Ryan Hill,
  Aroosa Ijaz, Theodor Isacsson, David Ittah, Soran Jahangiri, Prateek Jain,
  Edward Jiang, Ankit Khandelwal, Korbinian Kottmann, Robert~A. Lang, Christina
  Lee, Thomas Loke, Angus Lowe, Keri McKiernan, Johannes~Jakob Meyer, J.~A.
  {Monta{\~n}ez-Barrera}, Romain Moyard, Zeyue Niu, Lee~James O'Riordan, Steven
  Oud, Ashish Panigrahi, Chae-Yeun Park, Daniel Polatajko, Nicol{\'a}s Quesada,
  Chase Roberts, Nahum S{\'a}, Isidor Schoch, Borun Shi, Shuli Shu, Sukin Sim,
  Arshpreet Singh, Ingrid Strandberg, Jay Soni, Antal Sz{\'a}va, Slimane
  Thabet, Rodrigo~A. {Vargas-Hern{\'a}ndez}, Trevor Vincent, Nicola Vitucci,
  Maurice Weber, David Wierichs, Roeland Wiersema, Moritz Willmann, Vincent
  Wong, Shaoming Zhang, and Nathan Killoran.
\newblock {{PennyLane}}: {{Automatic}} differentiation of hybrid
  quantum-classical computations, July 2022.
\newblock URL \url{https://doi.org/10.48550/arXiv.1811.04968}.

\bibitem[Sun et~al.(2020)Sun, Zhang, Banerjee, Bao, Barbry, Blunt, Bogdanov,
  Booth, Chen, Cui, Eriksen, Gao, Guo, Hermann, Hermes, Koh, Koval, Lehtola,
  Li, Liu, Mardirossian, McClain, Motta, Mussard, Pham, Pulkin, Purwanto,
  Robinson, Ronca, Sayfutyarova, Scheurer, Schurkus, Smith, Sun, Sun, Upadhyay,
  Wagner, Wang, White, Whitfield, Williamson, Wouters, Yang, Yu, Zhu,
  Berkelbach, Sharma, Sokolov, and Chan]{sun2020}
Qiming Sun, Xing Zhang, Samragni Banerjee, Peng Bao, Marc Barbry, Nick~S.
  Blunt, Nikolay~A. Bogdanov, George~H. Booth, Jia Chen, Zhi-Hao Cui, Janus~J.
  Eriksen, Yang Gao, Sheng Guo, Jan Hermann, Matthew~R. Hermes, Kevin Koh,
  Peter Koval, Susi Lehtola, Zhendong Li, Junzi Liu, Narbe Mardirossian,
  James~D. McClain, Mario Motta, Bastien Mussard, Hung~Q. Pham, Artem Pulkin,
  Wirawan Purwanto, Paul~J. Robinson, Enrico Ronca, Elvira~R. Sayfutyarova,
  Maximilian Scheurer, Henry~F. Schurkus, James E.~T. Smith, Chong Sun,
  Shi-Ning Sun, Shiv Upadhyay, Lucas~K. Wagner, Xiao Wang, Alec White,
  James~Daniel Whitfield, Mark~J. Williamson, Sebastian Wouters, Jun Yang,
  Jason~M. Yu, Tianyu Zhu, Timothy~C. Berkelbach, Sandeep Sharma, Alexander~Yu.
  Sokolov, and Garnet Kin-Lic Chan.
\newblock Recent developments in the {{PySCF}} program package.
\newblock \emph{The Journal of Chemical Physics}, 153\penalty0 (2):\penalty0
  024109, July 2020.
\newblock ISSN 0021-9606.
\newblock \doi{10.1063/5.0006074}.

\bibitem[Huggins et~al.(2021)Huggins, McClean, Rubin, Jiang, Wiebe, Whaley, and
  Babbush]{huggins2021}
William~J. Huggins, Jarrod~R. McClean, Nicholas~C. Rubin, Zhang Jiang, Nathan
  Wiebe, K.~Birgitta Whaley, and Ryan Babbush.
\newblock Efficient and noise resilient measurements for quantum chemistry on
  near-term quantum computers.
\newblock \emph{npj Quantum Information}, 7\penalty0 (1):\penalty0 1--9,
  February 2021.
\newblock ISSN 2056-6387.
\newblock \doi{10.1038/s41534-020-00341-7}.

\bibitem[Zhao et~al.(2021)Zhao, Rubin, and Miyake]{zhao2021}
Andrew Zhao, Nicholas~C. Rubin, and Akimasa Miyake.
\newblock Fermionic partial tomography via classical shadows.
\newblock \emph{Physical Review Letters}, 127\penalty0 (11):\penalty0 110504,
  September 2021.
\newblock ISSN 0031-9007, 1079-7114.
\newblock \doi{10.1103/PhysRevLett.127.110504}.

\bibitem[Choi et~al.(2022)Choi, Yen, and Izmaylov]{choi2022}
Seonghoon Choi, Tzu-Ching Yen, and Artur~F. Izmaylov.
\newblock Improving quantum measurements by introducing "ghost" {{Pauli}}
  products.
\newblock \emph{Journal of Chemical Theory and Computation}, 18\penalty0
  (12):\penalty0 7394--7402, December 2022.
\newblock ISSN 1549-9618, 1549-9626.
\newblock \doi{10.1021/acs.jctc.2c00837}.

\bibitem[Gresch and Kliesch(2023)]{gresch2023}
Alexander Gresch and Martin Kliesch.
\newblock Guaranteed efficient energy estimation of quantum many-body
  {{Hamiltonians}} using {{ShadowGrouping}}, September 2023.
\newblock URL \url{https://doi.org/10.48550/arXiv.2301.03385}.

\bibitem[Koridon et~al.(2021)Koridon, Yalouz, Senjean, Buda, O'Brien, and
  Visscher]{koridon2021}
Emiel Koridon, Saad Yalouz, Bruno Senjean, Francesco Buda, Thomas~E. O'Brien,
  and Lucas Visscher.
\newblock Orbital transformations to reduce the 1-norm of the electronic
  structure hamiltonian for quantum computing applications.
\newblock \emph{Phys. Rev. Res.}, 3:\penalty0 033127, Aug 2021.
\newblock \doi{10.1103/PhysRevResearch.3.033127}.

\bibitem[Hohenstein et~al.(2022)Hohenstein, Oumarou, {Al-Saadon}, Anselmetti,
  Scheurer, Gogolin, and Parrish]{hohenstein2022}
Edward~G. Hohenstein, Oumarou Oumarou, Rachael {Al-Saadon}, Gian-Luca~R.
  Anselmetti, Maximilian Scheurer, Christian Gogolin, and Robert~M. Parrish.
\newblock Efficient {{Quantum Analytic Nuclear Gradients}} with {{Double
  Factorization}}, July 2022.
\newblock URL \url{https://doi.org/10.48550/arXiv.2207.13144}.

\bibitem[Wierichs et~al.(2022)Wierichs, Izaac, Wang, and
  Lin]{wierichs2022generalparameter}
David Wierichs, Josh Izaac, Cody Wang, and Cedric Yen-Yu Lin.
\newblock General parameter-shift rules for quantum gradients.
\newblock \emph{{Quantum}}, 6:\penalty0 677, March 2022.
\newblock ISSN 2521-327X.
\newblock \doi{10.22331/q-2022-03-30-677}.
\newblock URL \url{https://doi.org/10.22331/q-2022-03-30-677}.

\bibitem[Rubin et~al.(2018)Rubin, Babbush, and McClean]{Rubin2018}
Nicholas~C Rubin, Ryan Babbush, and Jarrod McClean.
\newblock Application of fermionic marginal constraints to hybrid quantum
  algorithms.
\newblock \emph{New Journal of Physics}, 20\penalty0 (5):\penalty0 053020, may
  2018.
\newblock \doi{10.1088/1367-2630/aab919}.
\newblock URL \url{https://dx.doi.org/10.1088/1367-2630/aab919}.

\bibitem[Stokes et~al.(2020)Stokes, Izaac, Killoran, and Carleo]{Stokes2020}
James Stokes, Josh Izaac, Nathan Killoran, and Giuseppe Carleo.
\newblock Quantum {N}atural {G}radient.
\newblock \emph{{Quantum}}, 4:\penalty0 269, May 2020.
\newblock ISSN 2521-327X.
\newblock \doi{10.22331/q-2020-05-25-269}.
\newblock URL \url{https://doi.org/10.22331/q-2020-05-25-269}.

\bibitem[Meyer(2021)]{Meyer2021}
Johannes~Jakob Meyer.
\newblock Fisher {I}nformation in {N}oisy {I}ntermediate-{S}cale {Q}uantum
  {A}pplications.
\newblock \emph{{Quantum}}, 5:\penalty0 539, September 2021.
\newblock ISSN 2521-327X.
\newblock \doi{10.22331/q-2021-09-09-539}.

\bibitem[Amari(1998)]{amari1998}
Shun-ichi Amari.
\newblock {Natural Gradient Works Efficiently in Learning}.
\newblock \emph{Neural Computation}, 10\penalty0 (2):\penalty0 251--276, 02
  1998.
\newblock ISSN 0899-7667.
\newblock \doi{10.1162/089976698300017746}.

\bibitem[Liang et~al.(2019)Liang, Poggio, Rakhlin, and Stokes]{liang2019}
Tengyuan Liang, Tomaso Poggio, Alexander Rakhlin, and James Stokes.
\newblock Fisher-{{Rao Metric}}, {{Geometry}}, and {{Complexity}} of {{Neural
  Networks}}, February 2019.
\newblock URL \url{https://doi.org/10.48550/arXiv.1711.01530}.

\bibitem[As\'oth et~al.(2016)As\'oth, Oroszl\'any, and P\'alyi]{asboth16}
J\'anos~K. As\'oth, L\'aszl\'o Oroszl\'any, and Andr\'as P\'alyi.
\newblock \emph{A short course on topological insulators: band structure and
  edge states in one and two dimensions}.
\newblock Springer, 2016.
\newblock ISBN 9783319256078 9783319256054.

\bibitem[Zak(1989)]{zak1989}
J.~Zak.
\newblock Berry's phase for energy bands in solids.
\newblock \emph{Phys. Rev. Lett.}, 62:\penalty0 2747--2750, Jun 1989.
\newblock \doi{10.1103/PhysRevLett.62.2747}.

\bibitem[Hatsugai(2006)]{hatsugai2006}
Yasuhiro Hatsugai.
\newblock Quantized berry phases as a local order parameter of a quantum
  liquid.
\newblock \emph{Journal of the Physical Society of Japan}, 75\penalty0
  (12):\penalty0 123601, 2006.
\newblock \doi{10.1143/JPSJ.75.123601}.

\bibitem[Fukui et~al.(2005)Fukui, Hatsugai, and Suzuki]{fukui2005}
Takahiro Fukui, Yasuhiro Hatsugai, and Hiroshi Suzuki.
\newblock Chern numbers in discretized brillouin zone: Efficient method of
  computing (spin) hall conductances.
\newblock \emph{Journal of the Physical Society of Japan}, 74\penalty0
  (6):\penalty0 1674--1677, 2005.
\newblock \doi{10.1143/JPSJ.74.1674}.

\bibitem[Chern(1946)]{chern1946}
Shiing-shen Chern.
\newblock Characteristic {{Classes}} of {{Hermitian Manifolds}}.
\newblock \emph{Annals of Mathematics}, 47\penalty0 (1):\penalty0 85--121,
  1946.
\newblock ISSN 0003-486X.
\newblock \doi{10.2307/1969037}.

\bibitem[Citro and Aidelsburger(2023)]{citro2023}
Roberta Citro and Monika Aidelsburger.
\newblock Thouless pumping and topology.
\newblock \emph{Nature Reviews Physics}, 5\penalty0 (2):\penalty0 87--101,
  January 2023.
\newblock ISSN 2522-5820.
\newblock \doi{10.1038/s42254-022-00545-0}.

\bibitem[Thouless(1960)]{thouless1960}
D.~J. Thouless.
\newblock Stability conditions and nuclear rotations in the {{Hartree-Fock}}
  theory.
\newblock \emph{Nuclear Physics}, 21:\penalty0 225--232, November 1960.
\newblock ISSN 0029-5582.
\newblock \doi{10.1016/0029-5582(60)90048-1}.

\end{thebibliography}

\clearpage
\onecolumngrid

\appendix

\section{Bounding overlaps by change in ansatz parameters} 
\label{app:parameter-change}

The variational Berry phase for a real ansatz state (as introduced in section \ref{sec:real-ansatz}) is resolved as the argument of the boundary term $\arg[\braket{\psi(\thetaest(0))}{\psi(\thetaest(1))}]$. 
To obtain a nontrivial Berry phase, the initial and final parameters $\thetaest(0)$ and $\thetaest(1)$ need to be far enough to allow $\braket{\psi(\thetaest(0))}{\psi(\thetaest(1))} = -1$.
This implies that the optimal parameters need to change enough along the parametrization of the path (for $t$ going from $0$ to $1$).
In this appendix, we translate this into a lower bound on the one-norm-distance between initial and final parameters $\lVert \thetaest(1) - \thetaest(0) \rVert_1$.
As a consequence, we also find a lower bound on how much error can be allowed without on the final parameter $\thetaest(1)$ compromising the Berry phase measurement.

We first state a lemma which will be useful in the proof:
\begin{lemma} \label{lem:unitary-triangle-inequality}
    Given two unitary operators $U, U'$, each decomposed as a product of $N\geq2$ unitary operators $U = U_{N-1} ... U_{1} U_{0}$, there holds the bound $\lVert U - U' \rVert \leq \sum_{j \in [N]} \lVert U_j - U'_j\rVert$.
\end{lemma}
We prove this by induction. For $N=2$, the proof is by the triangle inequality
\begin{align}
    \lVert U_1' U_0' - U_1 U_0 \rVert 
    &= 
    \lVert U_1' (U_0' - U_0) + (U_1' - U_1) U_0 \rVert 
    \\&\leq
    \lVert U_1' \rVert \lVert (U_0' - U_0) \rVert +  \lVert (U_1' - U_1) \rVert \lVert U_0 \rVert
    \\&=  
    \lVert (U_0' - U_0) \rVert +  \lVert (U_1' - U_1) \rVert.
\end{align}
The proof for $N+1$ can be similarly be reduced to the proof for $N$:
\begin{align}
    \lVert U'_N U'_{N-1} ... U'_0 - U_N U_{N-1} ... U_0 \rVert \leq 
    \lVert U'_N - U_N \rVert + \lVert U'_{N-1} ... U'_0 - U_{N-1} ... U_0 \rVert.
\end{align}

Let us consider the case of real Ansatz Eq.~\eqref{eq:real-unitary-ansatz}.
We remind $U_i = e^{A_i \theta_i}$ for antisymmetric real $A_i$.
We can then bound the overlap between ansatz states using Lemma~\ref{lem:unitary-triangle-inequality},
\begin{align}
    1-\braket{\psi(\bftheta)}{\psi(\bftheta')} 
    &=
    1 - \bra{\psi_0} U^\dag(\bftheta) U(\bftheta') \ket{\psi_0}
    \\&= 
    \bra{\psi_0} U^\dag(\bftheta) [U(\bftheta) - U(\bftheta')] \ket{\psi_0}
    \\&\leq 
    \lVert U(\bftheta) - U(\bftheta') \rVert
    \\&\leq 
    \sum_{j \in [N]} \lVert U_j(\theta_j) - U_j(\theta'_j) \rVert
    \\&=
    \sum_{j \in [N]} \lVert e^{A_j \theta_j} - e^{A_j \theta_j'} \rVert
    \\&=
    2 \sum_{j \in [N]} \left\lVert \sin\left[\frac{A_j}{2}(\theta_j - \theta_j')\right] \right\rVert
    \leq \sum_{j \in [N]} \lVert A_j \rVert |\theta_j - \theta_j'|
\end{align}

The difference of parametrers $|\theta_j - \theta_j'|$ is always rescaled by the respective $\lVert A_j \rVert$; we can interpret this by considering $\theta_j$ and $A_j$ as dimensionful quantities, with inverse dimension to each other.
We can always redefine units rescaling $\theta_j$ and $A_j$ -- without loss of generality we choose units for which $\lVert A_j \rVert = 1$ (the only assumption being the boundedness of $A_j$).
Under this choice,
\begin{equation}
    \braket{\psi(\thetaest(0))}{\psi(\thetaest(1))} = -1
    \implies 
    \lVert \thetaest(0) - \thetaest(1) \rVert_1 \geq 2
\end{equation}
Thus the parameters need to change (in one-norm) by at least 2 along the path to achieve the same state and nontivial Berry phase.
By the same reasoning, an error on the final parameters $\delta \thetaest(1)$ bounded by $\lVert\delta\thetaest(1)\rVert_1 < 1$ will not change the argument (i.e. the sign) of the overlap, thus allowing to resolve the correct Berry phase.

\section{Bounding the norm of energy derivatives}
\label{app:norm-of-derivatives}

Suppose we have a variational state parameterized by $\bftheta$
\begin{equation}
    \ket{\psi(\bftheta)} = U(\bftheta) \ket{0} = \prod_{k=0}^{n_p-1} U_k(\theta_k) V_k \ket{0}
\end{equation}
on a $n_p$-dimensional manifold (within a larger Hilbert space), where the product is assumed to be taken in unitary composition order (right to left).
Assume each $U_k$ is generated by anti-hermitian operators $i A_k$ with unit norm
\begin{equation}
    U_k(\theta) = e^{i A_k \theta} , \quad A_k = A_k^\dagger, \quad \lVert A_k \rVert = 1.
\end{equation}
Given $H$ is an observable of known norm $\lVert H \rVert$, define
\begin{equation}
    E(\bftheta) = \bra{\psi(\bftheta)} H \ket{\psi(\bftheta)}.
\end{equation}
Define the tensors of derivatives,
\begin{align}
    \mathcal{G}_j(\bftheta) &= \frac{\partial}{\partial \theta_j} E (\bftheta) \\
    \mathcal{H}_{jk}(\bftheta) &= \frac{\partial^2}{\partial \theta_j \partial \theta_k} E (\bftheta) \\
    \mathcal{T}_{jkl}(\bftheta) &= \frac{\partial^3}{\partial \theta_j \partial \theta_k \partial \theta_l} E (\bftheta) \\
\end{align}
our goal is to bound their (vector-induced) 2-norms $\lVert \mathcal{G} \rVert$, $\lVert \mathcal{H} \rVert$, $\lVert \mathcal{T} \rVert$.

We first notice that
\begin{equation}
    \frac{\partial}{\partial \theta_j} U(\theta) =
    \left( \prod_{k=j}^{n_p-1} U_k(\theta_k) V_k \right)  
    i A_j
    \left( \prod_{k=0}^{j-1} U_k(\theta_k) V_k \right)
    = i \tilde{A}_j U(\theta)
\end{equation}
where $\lVert\tilde{A}_j\rVert = \lVert A_j \rVert = 1$, as conjugation by a unitary preserves norm.
Using this, we can get expressions for the tensors of derivatives in terms of commutators of Hermitian operators of known norm
\begin{align}
    \mathcal{G}_j(\bftheta) &= i \langle \psi(\bftheta) | [H, \tilde{A}_j] | \psi(\bftheta) \rangle \\
    \mathcal{H}_{jk}(\bftheta) &= - \langle \psi(\bftheta) | [[H, \tilde{A}_j], \tilde{A}_k] | \psi(\bftheta) \rangle \\
    \mathcal{T}_{jkl}(\bftheta) &= -i \langle \psi(\bftheta) | [[[H, \tilde{A}_j], \tilde{A}_k], \tilde{A}_l] | \psi(\bftheta) \rangle
\end{align}
A trivial bound on this involves bounding each commutator by its norm, e.g.
\begin{align}
    \lVert \mathcal{G}(\bftheta) \rVert^2 
    &= \sum_{k=0}^{n_p+1} \lVert \langle \psi(\bftheta) | [H, \tilde{A}_j] | \psi(\bftheta) \rangle \rVert^2
    \leq 4 n_p \lVert H \rVert^2.
\end{align}
It is an open question wether we can improve on this bound.

We can get a similar result for the derivatives with respect to the orbital rotations, considering the reparametrization described in Sec.~\ref{sec:OOPQC-NR-step}.
We call $\ket\psi$ the PQC ansatz state in the full (active + core + virtual) space, padded with virtual (core) registers of qubits in the state $\ket{0}$ ($\ket{1}$).
We drop explicit dependence on $C$ and $\bftheta$, and we make explicit the differential rotation parameters $\kappa$. 
The cost function is then
\begin{equation}
    E(\kappa) = \bra{\psi} e^{\sum_{pq} \kappa_{pq} E_{pq}} H e^{-\sum_{rs} \kappa_{rs} E_{rs}} \ket{\psi}
    , \quad E_{pq} = c^\dag_{p,\uparrow} c_{q,\uparrow} + c^\dag_{p,\downarrow} c_{q,\downarrow},
\end{equation}
where $E_{pq}$ is the generator of a spin-adapted orbital rotation.
Its derivatives at $\kappa=0$ [note that the index pairs $(pq)$ are collected in one index for the purpose of rotating higher order derivatives] are easily calculated to be
\begin{align}
    \mathcal{G}_{(pq)}(\kappa = 0) &= \bra{\psi} [H, E_{pq}] \ket{\psi}, \\
    \mathcal{H}_{(pq), (rs)}(\kappa = 0) &= \bra{\psi} [[H, E_{pq}], E_{rs}] \ket{\psi},\\
    \mathcal{T}_{(pq), (rs), (tu)}(\kappa = 0) &= \bra{\psi} [[[H, E_{pq}], E_{rs}], E_{tu}] \ket{\psi}.
\end{align}
Observing that $\lVert E_{pq} \rVert = 2$ we obtain the same result as above (up to constant factors 2, 4, 8 respectively, coming from this norm).

\section{Analytical orbital gradient and Hessian}
\label{app:analyticOO}
In this section, we expand on the estimation of analytic orbital gradient [right block of the vector in Eq.~\eqref{eq:composite-gradient}] and orbital-orbital and orbital-circuit Hessian [bottom right and top right blocks of the matrix in Eq.~\eqref{eq:composite-hessian}, respectively]. 
We show how after the reparametrization $C \gets C \cdot e^{-\bfkappa}$, the $\bfkappa$-derivatives of the cost function $E(C \cdot e^{-\bfkappa}, \bftheta)$ can be expressed as a linear function of the 1- and 2-electron RDM, and the mixed Hessian $\nabla_\bfkappa \nabla_\bftheta E(C \cdot e^{-\bfkappa}, \bftheta)$ can be expressed in terms of $\bftheta$-derivatives of the same RDM. 

We first define the 1- and 2-electron reduced density matrix (RDM) in the spin-restricted formalism
\begin{align}
    \label{eq:one_rdm}
    \gamma_{pq}(\bftheta) &= \bra{\psi(\bftheta)}E_{pq}\ket{\psi(\bftheta)}\\
    \label{eq:two_rdm}
    \Gamma_{pqrs}(\bftheta) &= \bra{\psi(\bftheta)}e_{pqrs}\ket{\psi(\bftheta)},
\end{align}
where $E_{pq} = \sum_\sigma c_{p\sigma}^\dagger c_{q\sigma}$ and $e_{pqrs} = \sum_{\sigma \tau} c_{p\sigma}^\dagger c_{r\tau}^\dagger c_{s\tau} c_{q\sigma} = E_{pq}E_{rs} - \delta_{qr}E_{ps}$.
Here, $p,q,r,s$ are meant to be general indices (either occupied, active or virtual),
where the state $\ket{\psi(\bftheta)}$ is to be intended as padded by two registers of qubits in the $\ket{0}^{\otimes 2 N_\text{V}}$ ($\ket{1}^{\otimes 2 N_\text{O}}$) state for the virtual (occupied) orbitals.
In the molecular orbital basis defined by $C$ [orbitals in Eq.~\eqref{eq:parameterized-mo}], we can write the Hamiltonian as
\begin{align}
    H = \sum_{pq} h_{pq} E_{pq} + \frac{1}{2}\sum_{pqrs} g_{pqrs} e_{pqrs}
\end{align}
where $h_{pq}$ and $g_{pqrs}$ are the one- and two-electron integrals (with spatial orbital indices $p,q,r,s$ ordered according to the chemists' convention), and they implicitly depend on $C$ through the MOs.
The expectation value of the Hamiltonian can then be written as a contraction of the integrals with the RDM,
\begin{align}
    E(C,\bftheta) = \bra{\psi(\bftheta)} H(C) \ket{\psi(\bftheta)} = \sum_{pq} [h(C)]_{pq} \gamma_{pq} + \frac{1}{2}\sum_{pqrs} [g(C)]_{pqrs} \Gamma_{pqrs}
\end{align}
where we made explicit the dependence on $C$.

To derive analytical orbital rotation derivatives, we closely follow Ref.~\cite{helgaker2000}. 
We start by separating the dependence on $\kappa$ of the reparametrized cost function $E(C\cdot e^{-\kappa}, \bftheta)$, by using the equivalent state transformation formalism provided by Thouless theorem \cite{thouless1960}
\begin{align}
    E(C\cdot e^{-\kappa}, \bftheta) = 
    \bra{\psi(\bftheta)} H(C \cdot e^{-\kappa}) \ket{\psi(\bftheta)}
    =
    \bra{\psi(\bftheta)} e^{\hat{\kappa}} H(C) e^{-\hat{\kappa}} \ket{\psi(\bftheta)}
\end{align}
where $\hat{\kappa} = \sum_{pq} \kappa_{pq} E_{pq}$ is the operator that generates a unitary on the Hilbert state space equivalent to the orbital rotation.
We know that the rotations where $p, q$ are both virtual indices form a redundant subgroup, so we can freeze the corresponding $\kappa_{pq} = 0$; the same is true for $p, q$ both core space indices. 
In other terms, $\kappa_{pq} = 0$ if $p,q \in V$ or $p,q \in O$, with $V$ and $O$ the sets of virtual and core indices.
The remaining elements of $\kappa_{pq}$ satisfy $\kappa_{pq} = - \kappa_{qp}$.
We define a vector of unique non-redundant orbital rotation parameters 
\begin{equation}
    \bfkappa = \{\kappa_{pq}, \forall{p\in O \cup A}, \forall{q\in A \cup V}: q>p\},
\end{equation}
and we redefine the cost function with respect to this vector,
\begin{equation}
    E(C, \bfkappa, \bftheta) \equiv E(C\cdot e^{-\kappa}, \bftheta).
\end{equation}
We are interested in the derivative with respect to this vector; we can always switch from the matrix $\kappa$ to the unraveled vector of unique non-redundant parameters $\bfkappa$, and vice versa.
By comparing the Taylor series in $\kappa$ with the Baker-Campbell-Hausdorff expansion:
\begin{align}
    E(\bftheta, \bfkappa) = \bra{\psi(\bftheta)}H\ket{\psi(\bftheta)} + \bra{\psi(\bftheta)}[\hat{\kappa}, H] \ket{\psi(\bftheta)} + \frac{1}{2} \bra{\psi(\bftheta)}[\hat{\kappa}, [\hat{\kappa}, H]]\ket{\psi(\bftheta)} + \dots
\end{align}
One can readily verify that the analytical orbital derivatives at $\kappa_{pq} = 0$ are given by:
\begin{align}
    \label{eq:analytic_gradient_comm}
    [\nabla_{\bfkappa} E]_{pq} := \left.\frac{\partial E(\bftheta, \bfkappa)}{\partial \kappa_{pq}}\right|_{\kappa = 0} &= \bra{\psi(\bftheta)}[E^-_{pq}, H] \ket{\psi(\bftheta)} \\
    \label{eq:analytic_hessian_comm}
    [\nabla^2_{\bfkappa} E]_{pqrs} := \left.\frac{\partial^2 E(\bftheta, \bfkappa)}{\partial \kappa_{pq}\partial \kappa_{rs}}\right|_{\kappa = 0} &=
    \frac{1}{2} (1 + P_{pq,rs}) \bra{\psi(\bftheta)}[E^-_{pq}, [E^-_{rs}, H]]\ket{\psi(\bftheta)}
\end{align}
where $P_{pq,rs}$ permutes the pair of indices $pq$ with $rs$. The calculation of the commutators in Eq.~\eqref{eq:analytic_gradient_comm} and \eqref{eq:analytic_hessian_comm} can be found in common quantum chemistry textbooks \cite{helgaker2000}, and they all one- or two-body operators; thus their expectation value can be written as a linear form in the RDM ($\gamma$, $\Gamma$).
The gradient evaluates to
\begin{align}
    \label{eq:analytical_gradient}
    [\nabla_{\bfkappa} E]_{pq} = 2 (F_{pq}(\bftheta) - F_{qp}(\bftheta))
\end{align}
where $F$ is the generalized Fock matrix,
\begin{align}
    \label{eq:generalized_fock}
    F_{pq}(\bftheta) = \sum_{m} \gamma_{pm}(\bftheta) h_{qm} + \sum_{mnk} \Gamma_{pmnk}(\bftheta) g_{qmnk}
\end{align}
The Hessian evaluates to
\begin{align}
    \label{eq:analytical_hessian}
    [\nabla^2_{\bfkappa} E]_{pqrs} = (1-P_{pq})(1-P_{rs})\left[2\gamma_{pr}h_{qs} - (F_{pr} + F_{rp})\delta_{qs} + 2 Y_{pqrs}  \right],
\end{align}
where we introduced
\begin{align}
    Y_{pqrs} = \sum_{mn} \Gamma_{pmrn}g_{qmns} + \Gamma_{pmnr}g_{qmns} + \Gamma_{prmn}g_{qsmn}
\end{align}
and dropped the explicit dependence on $\bftheta$.
For the composite Hessian, we simply take the gradient of Eq.~\eqref{eq:analytical_gradient} with respect to $\bftheta$, by using the chain rule
\begin{align}
    [\nabla_{\bfkappa}\nabla_\bftheta E]_{pq}:= \left.\frac{\partial^2 E(\bftheta, \bfkappa)}{\partial \kappa_{pq}\partial \bftheta}\right|_{\kappa = 0} = 2 \left(\frac{\partial F_{pq}(\bftheta)}{\partial \bftheta} - \frac{\partial F_{qp}(\bftheta)}{\partial \bftheta}\right)
\end{align}
where
\begin{align}
    \frac{\partial F_{pq}(\bftheta)}{\partial \bftheta} = \sum_{m} \frac{\partial \gamma_{pm}(\bftheta)}{\partial \bftheta} h_{qm} + \sum_{mnk} \frac{\partial\Gamma_{pmnk}(\bftheta)}{\partial \bftheta} g_{qmnk}.
\end{align}
Thus, once we have the derivatives of the 1- and 2-RDM to sufficient precision, we can evaluate the orbital gradient, Hessian and composite Hessian analytically, recovering all terms in Eq.~\eqref{eq:composite-gradient} and Eq.~\eqref{eq:composite-hessian} without any additional quantum cost.

\section{Bounding the cumulative error due to Newton-Raphson updates}
\label{app:proof-NR-error-bound}

In this section, we prove that using a single Newton-Raphson (NR) parameter update per $\Delta t$-step is sufficient to achieve an error on the estimate of the minimizer scaling as $O(\Delta t^2)$ after any number of $\Delta t$-steps, as long as the cost function is strongly-convex at the minimum and $\Delta t$ is small enough.
First, we recall sufficient conditions for quadratic convergence of the Newton-Raphson step.
We then use these to bound the error of a single NR-step when the cost function is changed from $E(t,\bftheta) \to E(t+\Delta t,\bftheta)$.
We translate this into an upper bound on $\Delta t$ which guarantees the error stays bounded throughout the optimization path.
Finally, we show that we can allow a sufficiently small additive error on the Newton-Raphson update and retain the bounded error throughout the path.

\emph{Quadratic convergence of NR ---}
Consider a cost function $E(\bftheta)$ with gradient $\mathcal{G}_j=\frac{\partial E}{\partial \theta_j}$ and Hessian $\mathcal{H}_{jk}=\frac{\partial E}{\partial \theta_j \partial \theta_k}$, and an initial guess of a minimizer $\bftheta^{(0)}$.
The Newton-Raphson step prescribes the update $\bftheta^{(0)} \mapsto \bftheta^\text{NR} = \bftheta^{(0)} + d\bftheta^\text{NR}$ with $d\bftheta^\text{NR} = \mathcal{H}^{-1}(\bftheta^{(0)}) \mathcal{G}(\bftheta^{(0)})$.
Theorem 3.5 from Nocedal and Wright \cite{nocedal2006} gives sufficient conditions under which quadratic convergence of the NR update is guaranteed.
We simplify these conditions, and obtain the following
\begin{theorem} \label{thm:quadratic-convergence}
    Consider a cost function $E(\bftheta)$ 
    with Lipschitz-continuous Hessian $\lVert \mathcal{H}(\bftheta) -\mathcal{H}(\bftheta + \delta\bftheta)  \rVert \leq L \lVert \delta\bftheta \rVert$
    and a local minimizer $\thetaex$ with positive convexity $m := \lVert \mathcal{H}^{-1}(\thetaex) \rVert^{-1} > 0$.
    Given an initial guess $\bftheta^{(0)}$ which is close enough to the minimum, i.e.
    \begin{equation}
        \lVert \bftheta^{(0)} - \thetaex \rVert \leq \frac{m}{4L},
    \end{equation}
    the NR update will converge quadratically towards the minimum with
    \begin{equation}
        \lVert \bftheta^\text{NR} - \thetaex \rVert \leq \frac{L}{m} \lVert \bftheta^{(0)} - \thetaex \rVert^2.
    \end{equation}
\end{theorem}
To prove this, we only need to show that a strong convexity condition is satisfied within a $r$-ball centered in $\thetaex$ including all close-enough possible initial guesses ($r = \frac{m}{4L}$), i.e.
\begin{equation} \label{eq:half-convexity-region-requirement}
    \lVert \mathcal{H}^{-1}(\thetaex + \delta\bftheta) \rVert \leq 2 m^{-1} 
    ,\quad 
    \forall \lVert \delta\bftheta \rVert \leq \frac{m}{4L}.
\end{equation}
To prove this we expand $\mathcal{H}^{-1}(\thetaex + \delta\bftheta)$ using Taylor's theorem,
\begin{align}
    \exists 0<s<1 : \quad
    \mathcal{H}^{-1}(\thetaex + \delta\bftheta) 
    &= 
    \mathcal{H}^{-1}(\thetaex)
    + \delta\bftheta \cdot \frac{\partial \mathcal{H}^{-1}}{\partial \bftheta}(\thetaex + s \delta\bftheta)\label{eq:Taylor_first_step}
    \\
    \frac{\partial \mathcal{H}^{-1}}{\partial \bftheta} 
    &=
    -\mathcal{H}^{-1} \frac{\partial \mathcal{H}}{\partial \bftheta} \mathcal{H}^{-1}
    \\
    \lVert \mathcal{H}^{-1}(\thetaex + \delta\bftheta) \rVert
    & \leq
    \lVert \mathcal{H}^{-1}(\thetaex) \rVert + 
    \lVert \mathcal{H}^{-1}(\thetaex + s \delta\bftheta) \rVert^2 
    L \lVert \delta\bftheta \rVert
    \\
    & \leq
    m^{-1} + 
    \lVert \mathcal{H}^{-1}(\thetaex + s \delta\bftheta) \rVert^2 \frac{m}{4}
    ,
\end{align}
where we used the Lipschitz constant $L$ as a bound on the derivative of the Hessian.
This last condition holds if 
\begin{equation}\label{eq:Hessian_recursive}
    \lVert \mathcal{H}^{-1}(\thetaex + s \delta\bftheta) \rVert \leq 2 m^{-1}.
\end{equation}
As we can choose $s<1$ and the result is clearly true at $\delta\bftheta=0$, the result holds recursively.

\emph{Single NR update with a changing cost function ---}
We now consider a family of cost functions $E(t,\bftheta)$ continuously parameterized by $t$.
Suppose we have an approximation $\thetaest_t$ of the minimizer $\thetaex_t$ of $E(t, \bftheta)$, with error $\lVert \thetaest_t - \thetaex_t \rVert$.
In each step of our method (Algorithm~\ref{alg:main}), we shift the cost function $E(t,\bftheta) \to E(t+\Delta t,\bftheta)$ by $\Delta t$ and we use the current minimizer estimate $\thetaest_t$ as initial guess for the next step; $\bftheta^{(0)}_{t+\Delta t} = \thetaest_{t}$. 
We can bound the error of this initial guess by the triangle inequality,
\begin{equation}
    \lVert \bftheta^{(0)}_{t+\Delta t} - \thetaex_{t+\Delta t} \rVert 
    \leq
    \lVert \thetaest_t - \thetaex_t \rVert +
    \lVert \thetaex_{t+\Delta t} -  \thetaex_{t} \rVert.
\end{equation}
While the first term is the (given) error on the estimate, the second can be obtained by taking the total $t$-derivative of the minimum condition $\mathcal{G}(t, \thetaex(t)) = 0$, and applying Taylor's theorem
\begin{equation}
    \exists \tau \in [t, t+\Delta t]: \quad
    \lVert \thetaex_{t+\Delta t} -  \thetaex_{t} \rVert
    =
    \left\lVert \frac{d\thetaex_t}{d t}\Big|_{t=\tau} \right\rVert \Delta t
    =
    \lVert \mathcal{H}^{-1} (\tau, \thetaex_\tau) \dot{\boldsymbol{\mathcal{G}}} (\tau, \thetaex_\tau) \rVert \Delta t
\end{equation}
with $\dot{\boldsymbol{\mathcal{G}}} = \nabla \frac{\partial}{\partial t}E(\bftheta, t)$.

We now assume that the convexity at the minimum is bounded from below throughout the whole $t$-path by a constant $m \geq 0$,
\begin{equation}
    m(t, \thetaex_t) 
    := \lVert \mathcal{H}^{-1}(t, \thetaex_t) \rVert^{-1} > m
    \quad \forall t \in [0, 1],
\end{equation}
and that the gradient of the change is never larger than $\dot{\mathcal{G}}_\text{max}$.
(while the first assumption imposes the nontrivial condition of strong convexity at the minimum, the second is always granted for cost functions from a continuous family of bounded Hamiltonians).
We can then write
\begin{equation}
    \lVert \bftheta^{(0)}_{t+\Delta t} - \thetaex_{t+\Delta t} \rVert 
    \leq
    \lVert \thetaest_t - \thetaex_t \rVert +
    m^{-1} \dot{\mathcal{G}}_\text{max} \Delta t.
\end{equation}
To ensure this initial guess is within the quadratic convergence region of the NR step, we require 
\begin{equation}
    \lVert \thetaest_t - \thetaex_t \rVert +
    m^{-1} \dot{\mathcal{G}}_\text{max} \Delta t
    \leq
    \frac{m}{4L},
\end{equation}
choosing an $\alpha, \beta \in (0, 1]$ this condition can be written as
\begin{align}
    \Delta t &= \frac{m^2}{4L \dot{\mathcal{G}}_\text{max}} \alpha\beta, 
    \\ \label{eq:error-bound-before-delta-t-step}
    \lVert \thetaest_t - \thetaex_t \rVert &= (1-\alpha)\beta \frac{m}{4L}.
\end{align}
This allows to apply Theorem~\ref{thm:quadratic-convergence}, and bound the error after a single NR step,
\begin{equation} \label{eq:error-bound-after-delta-t-step}
    \lVert \thetaest_{t+\Delta t} - \thetaex_{t+\Delta t} \rVert
    \leq
    \frac{L}{m} 
    \left[
        \frac{\beta m}{4L}
    \right]^2 
    = \beta^2 \frac{m}{16 L}.
\end{equation}

\emph{Multiple steps ---}
We want to ensure the error on the minimizer estimate remains bounded for $t$ taking subsequent values is $[0, \Delta t, 2\,\Delta t, ..., 1]$, while taking a single NR step at a time.
We can do this by imposing the error after each step [Eq.~\eqref{eq:error-bound-after-delta-t-step}] is not larger than the error on the previous step estimate,
\begin{align}
    \beta^2 \frac{m}{16 L} \leq  (1-\alpha)\beta \frac{m}{4L}.
\end{align}
This is granted for any $\beta \in (0, 1]$ by choosing $\alpha = 1 - \frac{\beta}{4}$.
The maximum $\Delta t = \frac{3}{4}\frac{m^2}{4L \dot{\mathcal{G}}_\text{max}}$ is achieved by picking $\beta = 1$, and yields an error bounded by the constant $\frac{m}{16L}$.

\emph{Allowing an additive error ---}
To account for sampling noise, it is useful to consider an additive error of magnitude $\sigma_{\bftheta}$ on the estimate $\thetaest_t$ of the minimizer $\thetaex_t$ at each $t$-point, modifying Eq.~\eqref{eq:error-bound-before-delta-t-step} into
\begin{align}
    \lVert \thetaest_t - \thetaex_t \rVert &= (1-\alpha)\beta \frac{m}{4L} + \sigma_{\bftheta}.
\end{align}
This yields the condition
\begin{align}
    \frac{L}{m}\left[
        \frac{\beta m}{4L} + \sigma_{\bftheta}
    \right]^2  \leq  (1-\alpha)\beta \frac{m}{4L}.
\end{align}
If we define $\gamma$ by $\sigma_{\bftheta} = \frac{\gamma \beta m}{4 L}$, we can write 
\begin{align}
    (1 + \gamma)^2 \frac{m}{16L} \beta^2 
    \leq  
    (1-\alpha)\beta \frac{m}{4L},
\end{align}
which is saturated by $\beta = 4\frac{1-\alpha}{(1+\gamma)^2}$.
We then get
\begin{equation}
    \Delta t = \frac{m^2}{L \dot{\mathcal{G}}_\text{max}} \frac{\alpha(1-\alpha)}{(1+\gamma)^2}, 
\end{equation}
which is maximised (while keeping $\beta \leq 1$) by $\alpha = \max[\frac{1}{2}, 1  - \frac{(1+\gamma)^2}{4}]$.
The allowed additive noise is then
\begin{equation}
    \sigma_{\bftheta} = \frac{\gamma}{(1+\gamma)^2} \frac{m}{L} (1-\alpha) = 
     \frac{\gamma m}{2L} \min[
        \frac{1}{2}
    ,
        \frac{1}{(1+\gamma)^2}
    ],
\end{equation}
maximised for the choice $\gamma=\sqrt{2} - 1$ yielding
\begin{equation}
    \sigma_{\bftheta} = \frac{\sqrt{2}-1}{4} \frac{m}{L}
    ,\quad
    \Delta t = \frac{m^2}{8L \dot{\mathcal{G}}_\text{max}}.
\end{equation}

\section{Bounding the sampling cost}
\label{app:sampling-noise}

We call $\sigma_{\mathcal{G}}^2$ and $\sigma_{\mathcal{H}}^2$ the variances of each element of the gradient and Hessian respectively, due to sampling noise.
To compute the error on the parameter updates, we propagate these variances through the definition of the NR update Eq.~\eqref{eq:nr-step-definition}.
The first-order differential change (here denoted with $\delta$) of the NR update $d\bftheta^\text{NR}$ with respect to changes in the gradient and Hessian is
\begin{equation}
    \delta[d\bftheta^\text{NR}] = \mathcal{H}^{-1} \cdot [-\delta \boldsymbol{\mathcal{G}} + \delta \mathcal{H} \cdot d\bftheta^\text{NR}],
\end{equation}
where we use the ordered matrix-product notation, with vectors in boldface.
When $\delta \boldsymbol{\mathcal{G}}$ and $\delta \mathcal{H}$ are the random variables representing the errors on the gradient and Hessian, the expected mean square error on the NR update defining the norm of the covariance matrix
\begin{equation}
    \left\lVert \Var[d\bftheta^\text{NR}] \right\rVert
    :=
    \mathbb{E}\left[
        \lVert \delta[d\bftheta^\text{NR}] \rVert^2
    \right]
    =
    \mathbb{E}\left[
        \lVert \mathcal{H}^{-1} \cdot \delta \boldsymbol{\mathcal{G}} \rVert^2
    \right]
    +
    \mathbb{E}\left[
        \lVert \mathcal{H}^{-1} \cdot \delta\mathcal{H} \cdot d\bftheta^\text{NR} \rVert^2
    \right],
\end{equation}
where we used the zero-average property of $\delta \boldsymbol{\mathcal{G}}$ and $\delta \mathcal{H}$ to drop the expectation values of mixed terms.
This can further be bounded as
\begin{equation}
    \Var[d\bftheta^\text{NR}]
    \leq
    \lVert \mathcal{H}^{-1} \rVert^2 \mathbb{E}\left[
        \lVert \delta \boldsymbol{\mathcal{G}} \rVert^2
    \right]
    +
    \lVert \mathcal{H}^{-1} \rVert^2 \mathbb{E}\left[
        \lVert \delta\mathcal{H} \rVert^2
    \right] \lVert d\bftheta^\text{NR} \rVert^2.
\end{equation}
Assuming the same variance $\sigma^2_{\mathcal{G}}$ on each of the $n_p$ elements $\mathcal{G}_j$ of the gradient, we get 
\begin{equation}
    \mathbb{E}\left[ \lVert \delta \boldsymbol{\mathcal{G}} \rVert^2 \right] = n_p \, \sigma^2_{\mathcal{G}}.
\end{equation}
As the Hessian is a random real symmetric matrix with i.i.d.~elements, each with a variance $\sigma_\mathcal{H}$, we can invoke Wigner's semicircle law~\cite{wigner1955} to bound the spectral norm as 
\begin{equation}
    \mathbb{E}\left[ \lVert \delta \mathcal{H} \rVert^2 \right] \leq (\sqrt{n_p} \, \sigma_{\mathcal{H}})^2.
\end{equation}
Combining these with the strong convexity bound  $\lVert \mathcal{H}^{-1} \rVert \leq m^{-1}$, we get
\begin{equation} \label{eq:NR-update-variance}
    \Var[d\bftheta^\text{NR}]
    \leq
    m^{-2} \left[
        n_p \, \sigma^2_{\mathcal{G}}
        + n_p \, \sigma^2_{\mathcal{H}} \lVert d\bftheta^\text{NR} \rVert^2
    \right]
\end{equation}

The calculations in Appendix~\ref{app:proof-NR-error-bound} conclude that, for each $\Delta t$-step, we can afford an additive error on the NR update of at most $\sigma_{\bftheta} \leq \frac{\gamma}{4}\frac{m}{L}$ with $\gamma = \sqrt{2} - 1$.
Comparing this result to the variance just calculated, we can formulate the requirement on the variance
\begin{equation}
    m^{-2} n_p \left[
        \sigma^2_{\mathcal{G}}
        + \sigma^2_{\mathcal{H}} \lVert d\bftheta^\text{NR} \rVert^2
    \right] 
    \ll
    \frac{\gamma^2}{16}\frac{m^2}{L^2}.
\end{equation}
This gives conditions on the elementary sampling variances
\begin{align}
    \sigma^2_{\mathcal{G}}
    & \ll
    \frac{\gamma^2}{16}\frac{m^4}{L^2 n_p}
    \\
    \sigma^2_{\mathcal{H}} 
    & \ll
    \frac{\gamma^2}{16}\frac{m^4}{L^2 n_p \left\lVert d\bftheta^\text{NR} \right\rVert^2}
    <
    \frac{\gamma^2}{16}\frac{m^6}{L^2 n_p \lVert \dot{\mathcal{G}}_\text{max} \rVert^2 \Delta t^2}
\end{align}
To recast this bound in terms of variables of the problem, we use the following relations derived in Appendix~\ref{app:norm-of-derivatives}: $L = \max \lVert \mathcal{T} \rVert < n_p^{3/2} \lVert H \rVert$ (the norm of the third derivative tensor $\mathcal{T}$ is bounded by $n_p^{3/2}$ times by its infinity norm),  $\lVert \dot{\mathcal{G}}_\text{max} \rVert \Delta t < \sqrt{n_p} \lVert \frac{dH}{d t} \rVert \Delta t \approx \sqrt{n_p} 
\lVert H \rVert$ (same infinity norm bound).
Furthermore, we assume the convexity is larger than the ground state gap, $m > \Delta$; this holds if the ansatz approximates the ground state well enough, and changes in any ansatz parameter $\theta_k$ introduce a different excited state. 
Substituting these relations we obtain
\begin{align}
    \sigma^2_{\mathcal{G}}
    & \ll
    0.01 \frac{\Delta^4}{\lVert H \rVert^2 n_p^4},
    \\
    \sigma^2_{\mathcal{H}} 
    & \ll
    0.01\frac{\Delta^6}{\lVert H \rVert^4 n_p^5}.
\end{align}
The number of total required shots to sample the Hessian (gradient) for all the $N$ steps will thus scale as $\sigma^{-2}_{\mathcal{H}}  \Delta t^{-1}$ ($\sigma^{-2}_{\mathcal{G}} \Delta t^{-1}$). 
Picking the maximal $\Delta t = \frac{m^2}{8L \dot{\mathcal{G}}_\text{max}}$, and considering only the dominant term (relative to sampling the Hessian) we can write
\begin{equation}
    \text{\#shots} 
    \propto 
    n_p^7 \frac{\lVert H \rVert^7 \lVert \frac{dH}{d t}\rVert}{\Delta^8}
\end{equation}

\end{document}